\documentclass[utf8]{FrontiersinHarvard}

\usepackage{url,hyperref,lineno}
\hypersetup{
    colorlinks=true,
    linkcolor=blue,
    filecolor=magenta,      
    urlcolor=cyan,
    citecolor=purple,
    }
\usepackage[onehalfspacing]{setspace}
\usepackage{amsmath,amsfonts,amssymb}
\usepackage{graphicx}
\usepackage{setspace}
\usepackage{float}
\usepackage{caption}
\usepackage{subcaption}
\usepackage{soul}
\def\keyFont{\fontsize{8}{11}\helveticabold }
\def\firstAuthorLast{Marcotulli {et~al.}} 
\def\Authors{Lea Marcotulli\,$^{1,2,\dagger,*}$, Marco Ajello\,$^{3}$, Markus B\"ottcher$^{4}$, Paolo Coppi$^{5}$, Luigi Costamante$^{6}$, Laura Di Gesu$^{6}$, Manel Errando$^{7}$, Javier A. Garc\'ia$^{8,9}$, Andrea Gokus$^{7}$, Ioannis Liodakis$^{10,11}$, Greg Madejski$^{12}$, Kristin Madsen$^{8}$, Alberto Moretti$^{13}$, Riccardo Middei$^{6,14}$, Felicia McBride$^{15}$, Maria Petropoulou$^{16}$, Bindu Rani$^{17,18}$, Tullia Sbarrato$^{13}$, Daniel Stern$^{19}$, Georgios Vasilopoulos$^{16}$, Michael Zacharias$^{20,4}$, Haocheng Zhang$^{17, \ddagger }$ and the HEX-P Collaboration}

\def\Address{\tiny{$^{1}$Yale Center for Astronomy \& Astrophysics, 52 Hillhouse Avenue, New Haven, CT 06511, USA \\
$^{2}$Department of Physics, Yale University, P.O. Box 208120, New Haven, CT 06520, USA\\
$^{3}$Department of Physics and Astronomy, Clemson University, Kinard Lab of Physics, Clemson, SC 29634- 0978, USA\\
$^{4}$Centre for Space Research, North-West University, Potchefstroom 2520, South Africa\\
$^{5}$ Department of Astronomy, Yale University, P.O. 208101, New Haven, CT, 06520-8101, USA \\
$^{6}$ASI - Italian Space Agency, Via del Politecnico snc, 00133  Roma (RM), Italy \\
$^{7}$Department of Physics \& McDonnell Center for the Space Sciences, Washington University in St. Louis, One Brookings Drive, St. Louis, MO 63130, USA\\
$^{8}$X-ray Astrophysics Laboratory, NASA Goddard Space Flight Center, Greenbelt,
MD 20771, USA\\
$^{9}$Cahill Center for Astrophysics, California Institute of Technology, Pasadena, CA 91125, USA\\
$^{10}$Finnish Centre for Astronomy with ESO, 20014 University of Turku, Finland\\
$^{11}$NASA Marshall Space Flight Center, Huntsville, AL 35812, USA\\
$^{12}$SLAC National Accelerator Center and Kavli Institute for Particle Astrophysics and Cosmology, Stanford University, Stanford, CA 94305\\
$^{13}$INAF - Osservatorio Astronomico di Brera, via E.\ Bianchi 46, 23807 Merate (LC), Italy\\
$^{14}$INAF Osservatorio Astronomico di Roma, Via Frascati 33, 00078 Monte Porzio Catone (RM), Italy\\
$^{15}$Department of Physics and Astronomy, Bowdoin College, Brunswick, ME 04011, USA\\
$^{16}$Department of Physics, National and Kapodistrian University of Athens, University Campus Zografos, GR 15783, Athens, Greece\\
$^{17}$NASA Goddard Space Flight Center, Greenbelt, MD 20906, USA\\
$^{18}$University of Maryland, Baltimore County, Baltimore, MD 21250, USA\\
$^{19}$Jet Propulsion Laboratory, California Institute of Technology, Pasadena, CA 91109, USA\\
$^{20}$Landessternwarte, Universität Heidelberg, Königstuhl 12, 69117 Heidelberg, Germany\\
$\dagger$ NHFP Einstein Fellow\\
$\ddagger$ NASA Postdoc Program Fellow}}

\def\corrAuthor{Corresponding Author}

\def\corrEmail{lea.marcotulli@yale.edu}

\begin{document}
\onecolumn

\firstpage{1}

\title {The High Energy X-ray Probe (HEX-P): the most powerful jets through the lens of a superb X-ray eye} 

\author[\firstAuthorLast ]{\Authors} 
\address{} 
\correspondance{} 

\extraAuth{}

\maketitle
\begin{abstract}
A fraction of the active supermassive black holes at the centers of galaxies in our Universe are capable of launching extreme kiloparsec-long relativistic jets. These jets are known multiband (radio to $\gamma$-ray) and multimessenger (neutrino) emitters, and some of them have been monitored over several decades at all accessible wavelengths. However, many open questions remain unanswered about the processes powering these highly energetic phenomena. 
These jets intrinsically produce soft-to-hard X-ray emission that extends from $E\sim0.1\,\rm keV$ up to $E>100\,\rm keV$. 
Simultaneous broadband X-ray coverage, combined with excellent timing and imaging capabilities, is required to uncover the physics of jets. Indeed, truly simultaneous soft-to-hard X-ray coverage, in synergy with current and upcoming high-energy facilities (such as IXPE, COSI, CTAO, etc.) and neutrino detectors (e.g., IceCube), would enable us to disentangle the particle population responsible for the high-energy radiation from these jets. A sensitive hard X-ray survey ($F_{8-24\,\rm keV}<10^{-15}\,\rm erg~cm^{-2}~s^{-1}$) could unveil the bulk of their population in the early Universe. Acceleration and radiative processes responsible for the majority of their X-ray emission would be pinned down by microsecond timing capabilities at both soft and hard X-rays. Furthermore, imaging jet structures for the first time in the hard X-ray regime could unravel the origin of their high-energy emission. The proposed Probe-class mission concept High Energy X-ray Probe (HEX-P) combines all these required capabilities, making it the crucial next generation X-ray telescope in the multi-messenger, time-domain era. HEX-P will be the ideal mission to unravel the science behind the most powerful accelerators in the universe. 

\tiny
 \keyFont{ \section{Keywords:} blazar, high-energy astrophysics, X-ray mission, jets, supermassive black hole, multimessenger astrophysics, time-domain astrophysics} 
\end{abstract}

\section{Introduction}
\label{sec:intro}  
The high-energy ($E>1\,\rm keV$) extragalactic sky is filled with accreting supermassive black holes (known as active galactic nuclei, AGN, \citealp{Ajello_2008, Ajello_2009, Gilli_2013, Ueda_2014, Capelluti_2017, Ananna_2020}). These monstrous objects, reaching masses of billions or more Suns, are feeding from the gas that surrounds them, emitting copious amounts of radiation. A fraction 
of these black hole systems also powers relativistic jets from their core. When the jet is closely aligned to the Earth’s line of sight (viewing angles, $\theta_{\rm V}<5^{\circ}-10^{\circ}$), these objects are known as blazars; otherwise they are identified as radio galaxies \citep[e.g.,][]{urry_1995,Hovatta_2009,Mojave_xiv_2017}. Although continuous multi-band monitoring of these sources has been ongoing for more than two decades \citep[e.g.,][]{urry97,kataoka99,Komossa_2021, Weaver_2022, deJaeger_2023, Otero-Santo_2023}, many open questions remain regarding the nature of these extreme jets. 

A unique mission such as the High Energy X-ray Probe \citep[HEX-P;][]{Madsen23}, with its broadband X-ray coverage ($0.2-80\,\rm keV$), unprecedented sensitivity ($F_{0.5-2\,\rm keV}<10^{-16}\,\rm erg~cm^{-2}~s^{-1}$ and $F_{8-24\,\rm keV}<10^{-15}\,\rm erg~cm^{-2}~s^{-1}$ in a $1\,\rm Ms$ exposure, \citealp{Civano23}), good timing resolution ($\Delta t\sim2\,\rm \mu s$) across a wide range of energies and angular resolution in the hard X-ray band $3-4$ times better NuSTAR,
will enable us to answer several open science questions regarding blazars. HEX-P will allow us, for the first time, to: (1) spatially resolve extragalactic nearby jets in hard X-rays; (2) detect spectral signatures (e.g.~spectral curvatures, exponential cut-off) that will uniquely point us to the preferred high-energy emission mechanisms in jets, and, by extension, their composition; and (3) in synergy with the current and upcoming landscape of experiments (e.g., COSI, IXPE, CTAO, IceCube Gen 2), reveal the preferred particle acceleration and  emission mechanism of the brightest objects in the Universe, fulfilling the promises of multi-wavelength and multi-messenger science. In this paper, we explore some of the open issues in blazar science, and more broadly jetted AGN, and how HEX-P will revolutionize our understanding of these sources.

The peculiar jet orientation in blazars allows for extreme relativistic beaming, opening a window for studies of particle acceleration mechanisms and jet structure \citep{Madau_1987,Dermer_1995, Dondi_1995}. These sources are extremely bright multi-wavelength emitters (from radio up to TeV energies; \citealp[see][]{Madejski_2016, Blandforf_2019}), reaching 
bolometric luminosities of $>10^{48}\rm erg~s^{-1}$ \citep[e.g.][]{BZCAT_5, Ghisellini_2017, Paliya_2019, 4LAC-DR2}. Blazars are also likely multi-messenger sources, as they are able to accelerate protons and produce both high-energy neutrinos and gamma rays \citep[e.g.,][]{txs}.
Phenomenologically speaking, blazars have been divided into several subcategories. Objects with broad and strong emission lines (equivalent width, $\rm EW>5$\AA) in optical spectra are classified as Flat Spectrum Radio Quasars (FSRQs). When these lines are weak or absent, these sources are known as BL Lacertae objects \citep[BL Lacs, e.g.,][]{Schmidt_1968, Stein_1976}. 
Their electromagnetic broadband spectral energy distribution (SED) is characterized by two broad “humps”, a low-energy hump peaking in the IR--X-ray range, and a high-energy hump peaking in the {MeV}--$\gamma$-ray range.
It is now well understood that most of the radiation from radio to optical/X-ray frequencies of these jets is produced by synchrotron radiation of electrons moving at relativistic speeds
in magnetic fields in the jets. 
This scenario is further strengthened by the high polarization degree measured in these sources at radio to optical frequencies, and now also in X-rays for the two nearest blazars \citep[e.g.,][]{Jorstad_2007,Blinov_2021,Liodakis2022,Zhang_2023,Zobnina_2023}. 
Based on the location of their synchrotron spectral peak \citep{Abdo_2010}, these sources have later been classified as low- (LSP, $\nu_{\rm synch, peak}<10^{14.5}\,\rm Hz$), intermediate- (ISP, $10^{14.5}\,{\rm Hz}<\nu_{\rm synch, peak}<10^{15}\,{\rm Hz}$) and high-synchrotron peaked (HSP, $\nu_{\rm synch, peak}>10^{15}\,\rm Hz$) sources. FSRQs usually belong to the LSP class, being the most luminous of the class, detected up to the highest redshifts ($z>4$). BL Lacs instead span the whole range of classifications, are less luminous, and are typically found at $z<1$ \citep{Shen_2011,ajello14, Massaro_2016, Pena-Herazo_2021, Middei2022, Rajagopal_2023}.
However, the physical origin of the high-energy part of the SED is still a matter of debate. In the leptonic scenario \citep[e.g.,][]{Ghisellini_1989,Dermer_1992,Sikora_1994}, the X-ray to very high energy 
(VHE, $E > 100$~GeV) $\gamma$-ray emission is explained by inverse Compton scattering by the same population of electrons that produces the synchrotron emission. These electrons scatter the photon fields either within the jet structure (referred to as synchrotron self-Compton; SSC) or those external to the jet (for example the accretion disk, the broad-line region, or an infrared-emitting dust torus; this emission is referred to as external Compton; EC). 
As an alternative to the leptonic inverse-Compton interpretation, the high-energy emission could also be produced by ultra-relativistic protons through proton synchrotron emission and emission from electromagnetic cascades initiated by photo-pion production and Bethe-Heitler pair production processes \citep[e.g.,][]{Mannheim_1993,Mucke_2003}. 

Clues towards the jets' particle composition lie in the simultaneous characterization of the X-ray spectrum from soft X-rays 
($E \sim 1$~keV) to hard X-rays ($E \gtrsim 10$~keV).
This energy band is crucial for several blazar classes: it allows characterization of the synchrotron peak of the most extreme BL Lac sources \citep[e.g.,][]{costamante01}, pinpointing the high-energy peak of the most luminous MeV-peaked blazars \citep[e.g.,][]{2015ApJ...807..167T,2020ApJ...889..164M}, and characterization of the transition between the low- and high-energy humps of the SED \citep[e.g.,][]{Boettcher2013}, 
the shape of which is fundamental to unveil the nature of the particle population in the jet. Due to its coverage of the full X-ray energy range, HEX-P will be complementary for blazar science to many other recent or near-future 
experiments,
such as the recently launched Imaging X-ray Polarimetry Explorer (IXPE; $2-8\,\rm keV$, \citealp{IXPE}), the approved Compton Spectrometer and Imager (COSI; \citealp{COSI}), and the order-of-magnitude improvement in sensitivity to high-energy $\gamma$-rays above 100~GeV provided by the upcoming Cherenkov Telescope Array Observatory (CTAO; \citealp{CTA}). 

Another observational multi-band characteristic of blazars is their variability, with flaring timescales 
ranging from months down to minutes, 
flux enhancements of up to several orders of magnitude, and cross-correlated variability patterns (or lack thereof) observed between different energy bands. To explain this behavior, several particle acceleration mechanisms have been proposed to act in these jets, related to magnetic reconnection occurring inside the jet \citep[e.g.][]{giannios13,christie2019,Zhang2022}, 
or mildly relativistic shocks \citep{Rees_1978,Marscher_1985}. 
However, correlation studies between several energy bands are still inconclusive in diagnosing the production sites of flares (e.g., single zone versus multi-zone scenarios), the mechanisms of particle acceleration, and the particle species responsible for the flares in different wavelengths. Strictly simultaneous multi-wavelength observations and monitoring of flaring blazars must be performed to constrain jet physics. In particular, simultaneously characterizing the transition between the soft and hard X-rays in blazars belonging to different subclasses will provide unprecedented insight into these acceleration processes. 

Flaring can lead to a deeper understanding of the physics in AGN jets in terms of particle acceleration mechanisms and dissipative locations in the jet. Additionally, together with neutrino observations, flaring can be pivotal for understanding the particle content responsible for the photon emission. 
The association of the $\gamma$-ray flaring blazar TXS~0506+056 with the IceCube neutrino event IceCube-170922A was the first, and so far most convincing, evidence that blazars can accelerate protons to very high energies \citep{IceCube2018}. 
However, the lack of simultaneous broadband coverage from X-ray to MeV $\gamma$-rays has prevented us from constraining the relevant models sufficiently well to probe the transition between the synchrotron and high-energy emission part of the spectrum, where signatures of electromagnetic cascades associated with the neutrino-producing photo-hadronic interactions should emerge \citep{Murase2018}. In particular, studies of a previous neutrino flare potentially linked to TXS~0506+056 show that the complete lack of sensitive simultaneous broad X-ray spectral and temporal coverage significantly hampered any conclusion on the emission mechanism \citep[e.g.][]{Reimer19, Zhang2020, Petropoulou_2020}. 
Simultaneous hard X-rays ($E>10\,\rm keV$) would be able to probe the transition regime between the low-energy and high-energy peak, and the evolution of the spectral shape at different epochs (pre/post neutrino event) hence removing degeneracies on parameters in the tested models. Moreover, in the context of one-zone hybrid leptonic scenario, \citet{Petropoulou_2020} found that the maximal neutrino flux is strongly correlated with the X-ray flux in the case of the source, which underscores the importance of X-ray measurements for constraining the blazar neutrino output.
Simultaneous broadband coverage of neutrino alerts in the direction of blazar sources could be the tipping point for jet models as well as confirming AGN as one of the sources of high-energy cosmic rays. 

In terms of population studies, there appears to exist an anti-correlation between the synchrotron peak position and the intrinsic bolometric luminosity of blazars, the so-called blazar sequence \citep[see][]{Fossati_1998,Ghisellini_1998,Ghisellini_2017}. If real, this would imply that more luminous sources (found at the highest redshifts) have higher accretion rates ($L_{\rm disk}/L_{\rm Edd}\gtrsim 10^{-2}$), hence host more radiatively efficient accretion disks 
than less luminous (and lower redshift) ones. Opponents of the sequence, however, argue that this is a mere selection effect \citep[e.g.,][]{Nieppola_2008,Giommi_2012,Keenan_2021}. 
A mission with high sensitivity in the broad X-ray band such as HEX-P will be key to narrowing down this issue. On the one hand, HEX-P's superior sensitivity will allow us to detect tens to hundreds of sources beyond $z \gtrsim 4$, as predicted by current X-ray blazar evolutionary models (the so-called MeV blazars; \citealp[see,][]{Marcotulli_2022}).  Multi-band coverage will enable us to pinpoint the location of the high-energy peak 
and the intrinsic jet luminosity (probing the low-synchrotron-peak-frequency end of the blazar sequence). On the other hand, HEX-P has the potential to discover a handful of MeV-peaked BL Lac sources in the local universe. Predicted by the blazar sequence, these jets may have been missed by current surveys due to being too faint or overwhelmed by galaxy emission. If found by HEX-P, they would confirm that the blazar sequence is real and clarify its physical origin. 

Finally, jets of active galaxies have been observed to extend up to hundreds of kiloparsecs into the intergalactic medium. 
They are spatially best resolved in the radio band, using very-long-baseline interferometry (VLBI) observations, where low-energy synchrotron emission can be observed along the entire jet \citep[e.g.][]{Linfield_1981,Boccardi_2016}. Moreover, many jets also show extended components (straight jets or hotspots along the jet flow) in various other energy bands including the X-ray domain \citep{Chartas_2000,Harris_2006,Meyer_2018, Marshall_2018}, even up to the TeV energy band \citep[e.g.][]{2012ApJ...746..151A,HESS20,2020ApJ...896...41A}. However, the origin of the X-ray (and high-energy) emission is still a matter of debate. The most commonly invoked scenarios are inverse-Compton scattering of cosmic microwave background (CMB) photons or synchrotron emission of highly relativistic electrons \citep[e.g.,][]{Stawarz_2003,Atoyan_2004,ZachariasWagner16,meyer+17,Roychowdhury+22}. Observations with a mission such as HEX-P will be crucial for resolving, for the very first time in the hard X-ray band beyond $10\,$keV, nearby bright jets and disentangling these scenarios.

The paper is organized as follows: Section~\ref{sec:mission} describes the HEX-P mission design; Section~\ref{sec:simulations} details the setup for the simulations performed throughout the paper; Section~\ref{sec:multimessenger} addresses the contribution of HEX-P to the multi-messenger and time domain astrophysics of blazars; Section~\ref{sec:source_classes} highlights the prospects for different blazar classes; Section~\ref{sec:jet_resolve} showcases the importance of HEX-P in resolving jets in the hard X-ray band; and Section~\ref{sec:concl} summarizes the main science cases that HEX-P will be able to address with respect to the most powerful jets in the Universe.

\section{Mission Design} \label{sec:mission}

HEX-P \citep{Madsen23} is a Probe-class mission concept that offers sensitive broad-band coverage ($0.2-80\,\rm keV$) of the X-ray spectrum with exceptional spectral, timing, and angular capabilities. It features two high-energy telescopes (HETs) that focus hard X-rays, and one low-energy telescope (LET) that focuses lower energy X-rays.

The LET consists of a segmented mirror assembly coated with iridium on monocrystalline silicon that achieves a half-power diameter of $3.5''$, and a low-energy DEPFET detector, of the same type as the Wide Field Imager \citep[WFI;][]{wfi} onboard Athena \citep{Nandra13}. It has 512 $\times$ 512 pixels that cover a field of view of $11.3'\times 11.3'$. It has an effective passband of $0.2-25\rm \,keV$, and a full frame readout time of 2\,ms, which can be operated in a 128 and 64 channel window mode for higher count-rates to mitigate pile-up and faster readout. Pile-up effects remain below an acceptable limit of $\sim 1\%$ for a flux up to $\sim 100$\,mCrab ($2.4\times10^{-9}\,\rm erg~cm^{-2}~s^{-1}$, in the $2-10\,\rm keV$) in the smallest window configuration. Excising the core of the PSF, a common practice in X-ray astronomy, will allow for observations of brighter sources, with a  typical loss of up to $\sim 60\%$ of the total photon counts.

The HET consists of two co-aligned telescopes and detector modules. The optics are made of Ni-electroformed full shell mirror substrates, leveraging the experience of XMM-Newton \citep{Jansen2001}, and coated with Pt/C and W/Si 
multilayers for an effective passband of $2-80\,\rm keV$, achieving a half-power diameter $3-4$ times better than that of NuSTAR. The high-energy detectors are of the same type as flown on NuSTAR 
\citep{Harrison2013}, and they consist of 16 CZT sensors per focal plane, tiled 4$\times$4, for a total of 128$\times$128 pixels spanning a field of view slightly larger than for the LET, of $13.4^\prime\times13.4^\prime$.


\section{Simulations}\label{sec:simulations}
All the simulations presented here were produced with a set of response files that represent the observatory performance based on current best estimates as of Spring 2023 \citep[see][]{Madsen23}. The effective area is derived from ray-tracing calculations for the mirror design including obscuration by all known structures. The detector responses are based on simulations performed by the respective hardware groups, with an optical blocking filter for the LET and a Beryllium window and thermal insulation for the HET. The LET background was derived from a GEANT4 simulation \citep{Eraerds2021} of the WFI instrument, and the HET background was derived from a GEANT4 simulation of the NuSTAR instrument; both backgrounds 
assume HEX-P's Lagrange point L1 orbit.

The spectral simulations were performed in {\sc xspec} \citep{xspec} via the `fakeit' command with V07\footnote{https://hexp.org/response-files/} response files, selecting an 80\% PSF correction.
The data are assumed to be extracted from a 10 arcsec region, with the anticipated background for the telescope being positioned at L1. 
The images and light curves were simulated using the SIXTE toolkit \citep{Dauser_2019}, using the same response files as the spectral simulation. 

The differential sensitivity curves shown throughout the paper are obtained by calculating for each energy bin (LET: $\Delta\rm E=0.01\,\rm keV$, HET: $\Delta\rm E=0.04\,\rm keV$) the minimum detectable flux for a signal-to-noise ratio (SNR) of 5 in exposures of $100\,\rm ks$ and $1\,\rm Ms$.

\section{Time domain and Multi-messenger Astrophysics}\label{sec:multimessenger}

\subsection{Leptonic vs. Hadronic Populations in Blazar Jets}
\label{sec:modeling_lep_hadr}\label{sec:particles}

\subsubsection{Probing the relativistic leptonic population in blazar jets}\label{sec:modeling_EC_SSC}\label{sec:lep_mod}

\begin{figure*}
\centering
\includegraphics[width=0.8\textwidth]{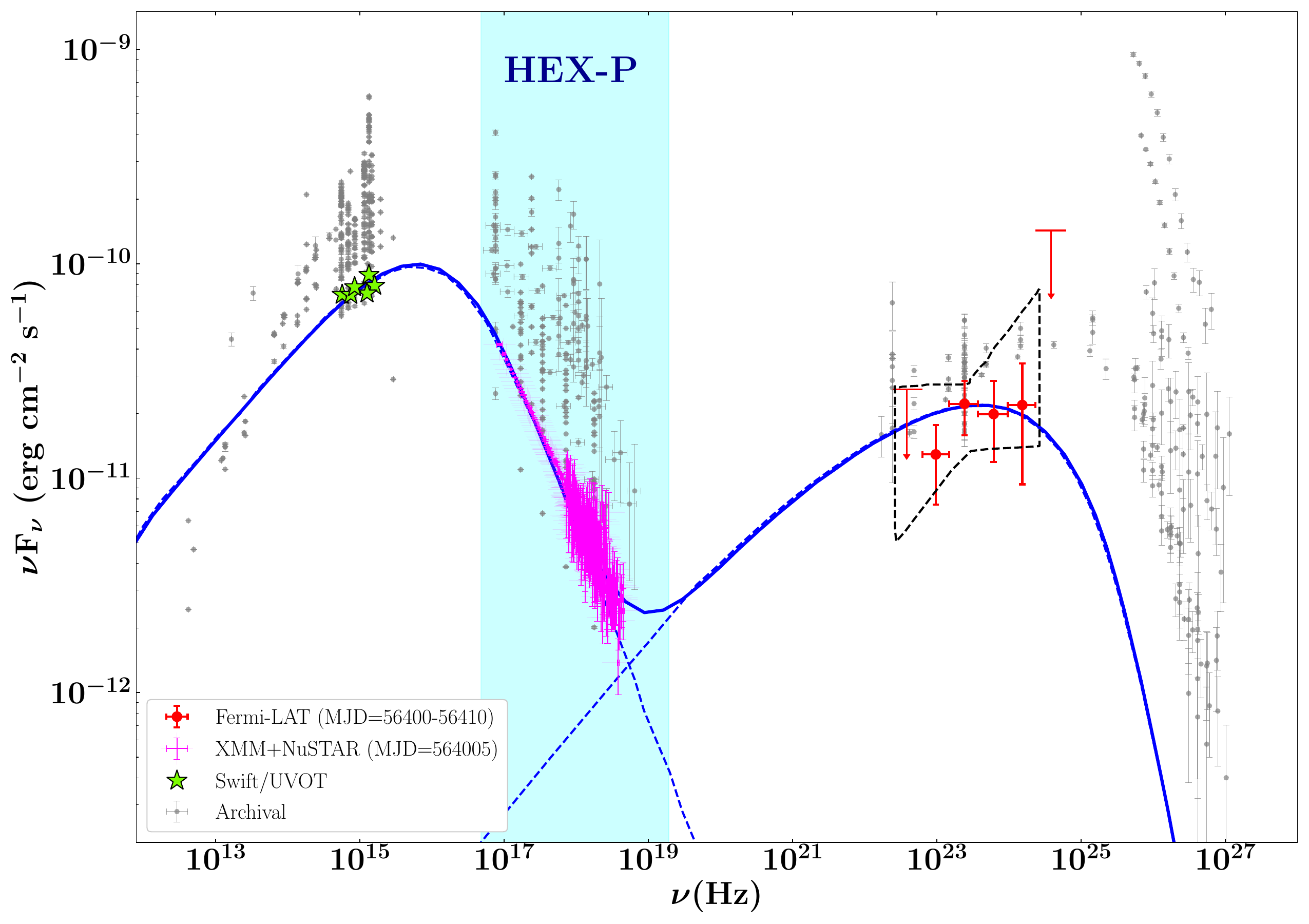}
\caption{{\bf Left:} Broadband SED of the HSP blazar PKS 2155-304 adapted from \citet{Madejski2016}. In gray, we show the multi-epoch multi-band archival data for the source, as an example of the extreme multi-band variability of blazar sources. The low-state was characterized by \citet{Madejski2016}, collecting contemporaneous data from: NuSTAR and XMM-Newton-pn (MJD=56405; magenta points), Swift-UVOT (fluxes corrected for Galactic reddening; green stars), and Fermi-LAT obtained for 10 days centered on the NuSTAR observation (data points in red; broad-band spectral fit “butterfly” in black).  Also shown is a basic SED model including synchrotron and SSC components (blue).  As mentioned in the text, the X-ray data occupy the ``valley'' in the SED.
}
\label{fig:pks2155_sed}
\end{figure*}

As discussed in the Section \ref{sec:intro}, the location of the two peaks of the blazar's broadband SED generally depends on its apparent luminosity \citep[e.g.,][]{Ghisellini_2017}. 
For HSP sources, the flux in the spectral regime corresponding to the X-ray energy ($0.1-100\,\rm keV$) is typically lower than that of the peaks. Thus in an SED plot $(log (E \times F(E))$ vs. $log (E)$, it appears as a ``valley'' (although note that the peak of the synchrotron component of HSPs and some LSPs and ISPs during flares can extend to the X-ray bands). Figure~\ref{fig:pks2155_sed} shows the broadband SED for the HSP blazar PKS 2155-304 (in a non-flaring state) from \citet{Madejski2016}, illustrating this ``valley". Recent measurements by the IXPE satellite have revealed that the soft X-ray ($0.2-8\,\rm keV$) emission of such sources is also polarized \citep{Liodakis2022}, confirming the synchrotron process as the dominant emission mechanism for the entire low-energy ``hump" in blazar SEDs.
This suggests that hard X-ray studies probe emission by the most energetic particles in the jet, since in the synchrotron process, the radiated photon energy
goes as the square of the energy of the radiating particle. 
Given the weak $E \times F(E)$ flux in the X-ray band, current instruments cannot accurately measure the spectral cutoff of the synchrotron tail during non-flaring states.   
With broadband coverage in both soft and hard X-ray energies, HEX-P will be able to accurately measure the spectral shape of the synchrotron tail, setting important constraints on the acceleration process of the radiating particles. Moreover, in the leptonic scenario, VHE $\gamma$-ray photons are produced by the same electrons that generate the X-ray synchrotron emission. Thus, HEX-P will probe the true electron distribution generating (by the inverse Compton process) 
the intrinsic VHE spectrum of HSPs, as well as flaring LSPs and ISPs that extend to X-rays. These constraints on the shape of the particle population are crucial as VHE photons from blazars are attenuated by $\gamma$-$\gamma$ interactions with low-energy photons associated with the extragalactic background light (EBL). In synergy with CTAO, this can provide unprecedentedly strong constraints on the EBL \citep[e.g.,][]{Desai_2019}. 

In the case of LSP sources, the X-ray band already probes the onset of the high-energy SED ``hump". 
In the context of synchro-Compton models for relativistic jets, this part of the spectrum is produced by the lowest-energy
part of the distribution of radiating particles.  
In these sources, therefore, HEX-P will probe the low-energy tail of the particle distribution, possibly revealing the presence of low-energy cutoffs thanks to the long arm provided by broad-band sensitivity out to 80 keV in a single observation \citep[e.g.][]{fabian_2001,tavecchio_2007,sbarrato_2016}.
The broad-band capability of HEX-P is also important in ISPs sources, where the X-ray band corresponds to the  cross-over between the synchrotron and inverse Compton peaks, i.e., the ``valley" floor \citep[see e.g., ON\,231 and TXS~0506+056,][]{on231,txs}, and the hard-X-ray range provides constraints on  
the slope of the low-energy electrons.
In both cases, given the observed photon spectrum, it is possible to infer the distribution of energies of radiating particles, and
for relativistic jets, the low-energy radiating particles are the most numerous.  This, in turn, provides good estimates of the total particle content of the jet, and
strong constraints on the total kinetic power of the jet \citep{Madejski2016}, as well as its electron-positron pair content.  
Thus, for high-luminosity sources, hard X-ray observations by HEX-P will provide
estimates of the jet power.  We note that, in principle, one could study the low-energy end of particle distributions by examining the lowest energy part of the
low-energy ``hump''. This, however, is not possible, because at low radio frequencies, where those low-energy particles are radiating via the synchrotron
process, the emission is self-absorbed, and thus the observed emission is optically thick, preventing robust measurements of the total content of low-energy particles in the jet.

\subsubsection{Probing the relativistic hadronic population in blazar jets}
\label{sec:modeling_hadro_lepto}

The high-energy blazar spectral component may consist of one or multiple radiation processes. In the leptonic model, the high-energy component may have contributions from both SSC and EC, while in the hadronic model, the hadronic cascading component can be considerable in the X-ray to MeV $\gamma$-ray range. In either case, the hard X-rays often mark the transition from one radiation process to another, which usually manifests itself through a change in the spectral shape \citep[e.g.,][]{Boettcher2013, Petropoulou_2015, Gao_2017}. But if the high-energy component is dominated by only one radiation mechanism (such as SSC), we expect a smooth, power-law like, hard X-ray spectrum. With its unprecedented spectral resolution and broadband coverage, HEX-P will be able to identify the presence of more than one radiation mechanism at X-ray energies.

To illustrate the capabilities of HEX-P in identifying spectral breaks due to the transition from leptonic to hadronic emission components, we choose Ap Librae, an intriguing LSP at redshift $z \simeq 0.049$ \citep{Disney_1974} that has been
been detected at VHE \citep{2015A&A...573A..31H}. It is the only TeV BL Lac object with a detected kpc-scale X-ray jet \citep{kaufmann+2013}. VHE data in combination with the
X-ray and high-energy (HE, $0.1-300$~GeV) $\gamma$-ray observations with Fermi-LAT, revealed an unexpected broad high-energy hump, spanning more than nine orders of magnitude in energy (from X-rays to TeV $\gamma$-rays). \cite{Petropoulou_2017} demonstrated that the superposition of different
emission components related to photohadronic interactions in the blazar core can explain the broad high-energy component of Ap Librae without invoking external radiation fields or particle acceleration at kpc scales \citep[for large-scale jet scenarios, see Section~\ref{sec:jet_resolve} and, e.g.,][]{ZachariasWagner16, Roychowdhury+22}. 

\begin{figure*}
    \includegraphics[width=0.49\textwidth]{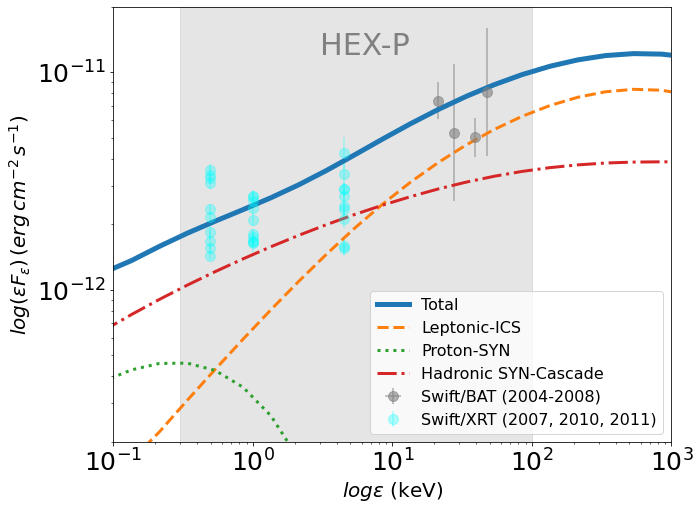}
        \includegraphics[width=0.47\textwidth]{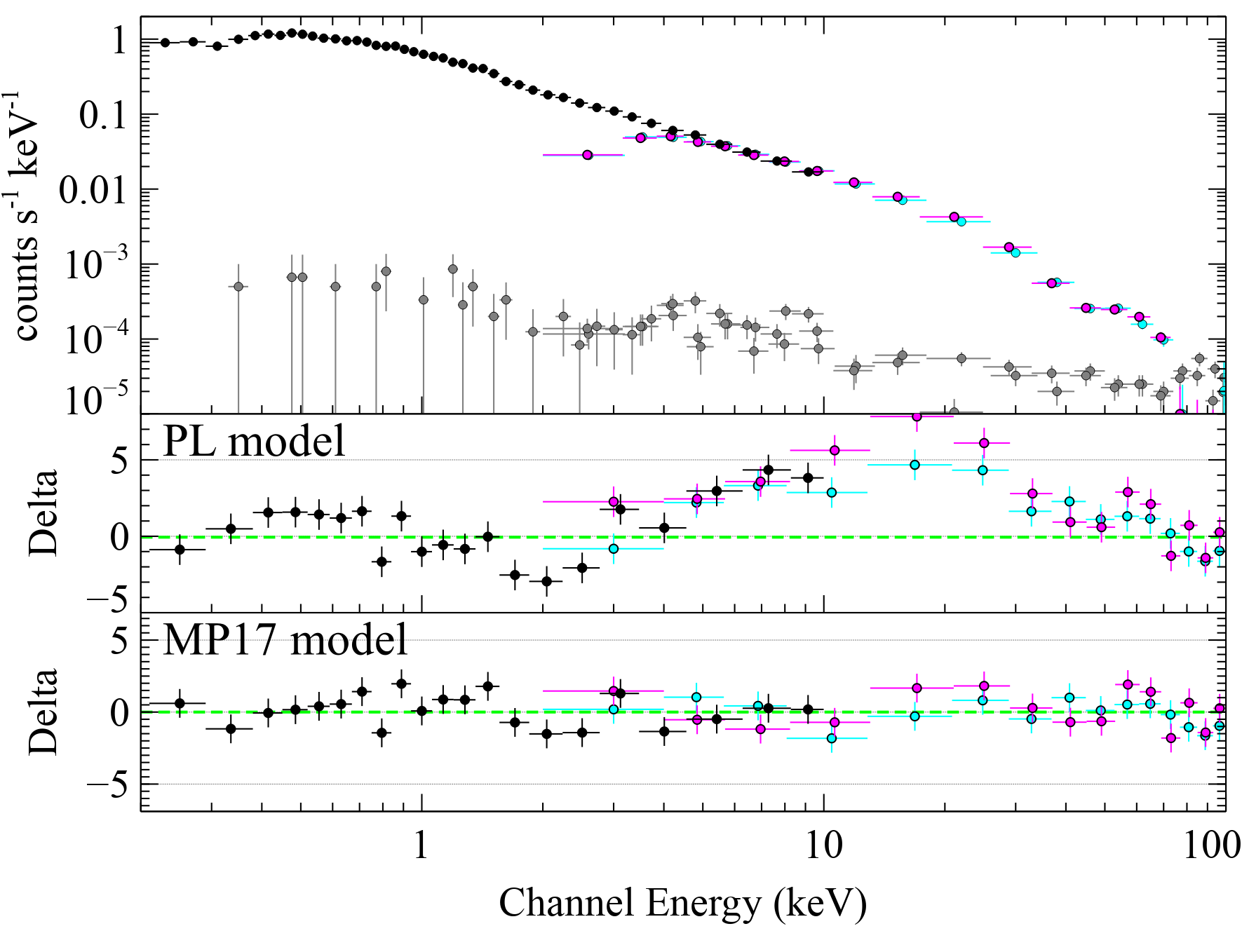}
    \caption{\textit{Left panel:} Theoretical spectrum for the core emission of Ap Librae from \cite{Petropoulou_2017} showing hadronic and leptonic contributions in the 0.2 - 100 keV range. Colored markers show non-simultaneous observations in soft and hard X-rays obtained with Swift/XRT and Swift/BAT, respectively. A smooth spectral break is evident at $\sim 2$~keV due to the onset of leptonic inverse Compton scatter (ICS) over the hadronic synchrotron (SYN) cascade emission. \textit{Right panel:} Simulated spectrum of Ap Librae for 50 ks exposure with HEX-P based on the theoretical spectrum shown on the left (top); background contribution is shown with gray shaded points. LET data are shown in black. Magenta and cyan colors correspond to the two high-energy detectors. Residual plots (i.e. (data-model)/error) of a power-law fit (middle) and a fit with a tabulated leptohadronic model (bottom) to the simulated spectrum. The residuals indicate that the power law model is disfavored.}
    \label{fig:aplib}
\end{figure*}

We adopt Model A from \cite{Petropoulou_2017} (see left panel in Figure~\ref{fig:aplib}) to simulate spectra for HEX-P, assuming a 50~ks exposure. The model predicts a broadband X-ray spectrum with leptonic and hadronic contributions in the energy range probed by HEX-P. More specifically, inverse Compton scattered (ICS) emission by accelerated electrons in the jet contributes mostly at hard X-rays ($>10$~keV), while synchrotron emission from secondary leptons produced in a hadronic-initiated cascade is dominating in soft X-rays.  The simulation results are presented in the right panel of  Figure~\ref{fig:aplib}. The residuals of a power-law fit to the broadband spectrum have some structure which is improved when the spectrum is modeled with a tabulated leptohadronic model. In the latter case the statistical improvement is of the order of 100 using $\chi^2$ or c-statistics for 5800 degrees of freedom (unbinned data). Moreover, the residuals at 2 keV, where the spectral break is expected, are reduced in the leptohadronic model. 
We caution, however, that similar spectral breaks may also occur in a purely leptonic scenario, at the transition from SSC to EC emission, if a sufficiently dense external radiation field exists. 

\begin{figure}
\centering
\includegraphics[width=0.4\textwidth]{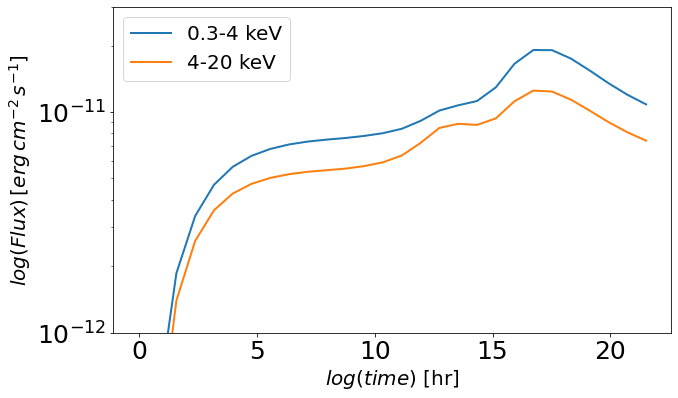}
\includegraphics[width=0.55\textwidth]{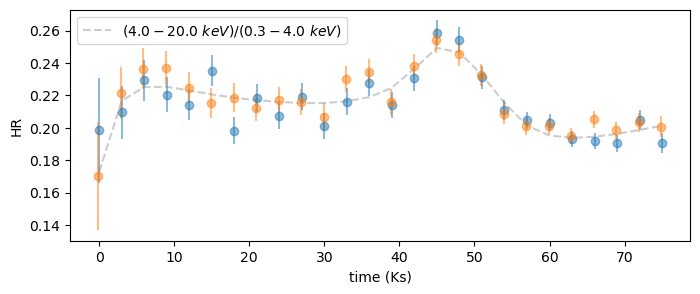}
\caption{\textit{Left panel:} Theoretical soft (0.3-4 keV) and hard (4-20 keV) X-ray light curves of a hypothetical flare from Ap~Librae with a total duration of 21~hr. \textit{Right panel:} Evolution of the hardness ratio (HR) during the flare. The HR evolution based on the tabulated model is shown with dashed line, while the results of two simulations are overplotted with colored markers. Each point is obtained assuming a 3~ks observation with HEX-P.}
\label{fig:aplib-HR}
\end{figure}

To demonstrate also the capabilities of HEX-P in detecting spectral changes during the course of an X-ray flare, we simulate a $\sim 1$-day long flare based on Model A of \cite{Petropoulou_2017} by introducing variations in the injection rate of accelerated electrons and protons in the blazar core. The injection rate of both particle species is modeled with a Lorentzian as a function of time. The applied changes in the proton injection rate lag behind the variations in the electron injection rate by $\sim 3$~hr. As a result, we expect to see first an increase in the hard X-ray flux, followed by an increase in softer X-rays where the hadronic cascade emission dominates (see left panel in Figure~\ref{fig:aplib}). The theoretical light curves of the flare in soft and hard X-rays are presented in the left panel of Figure~\ref{fig:aplib-HR}. We simulate 27 HEX-P spectra, 3~ks exposures each, and compute the hardness ratio (HR), defined as the ratio of counts in the 4-20 keV and 0.3-4 keV ranges. In Figure~\ref{fig:aplib-HR} we show the evolution of HR with time for two different simulations (blue and orange markers) that illustrate the statistical variance of the HR. Both simulations recover the expected changes in the HR shown by the dashed line.

\subsection{Multimessenger Follow-up:  Cosmics Rays and Neutrinos}
\label{sec:mm}
Cosmic rays were first detected at the beginning of the 20th century
\citep{Wulf1909, Pacini1912,Hess}, yet, more than 100 years later, the sources of the
high and ultra-high energy cosmic rays remain unclear. Since protons are
deflected in magnetic fields, they are not reliable tracers of their
origin. Neutrinos that are produced in proton-proton and proton-photon interactions travel in a
straight path, but are much harder to detect.

Blazars and AGN have long been suspected to be neutrino emitters. Material in the accretion disk and in the vicinity of the supermassive black hole includes partially or fully ionized gas, which is available for injection into the relativistic outflows. Protons in the jet can interact with photons from a stationary photon field (e.g., thermal UV emission from the disk), and produce neutral and charged pions. Charged pions decay into several products including several neutrinos, while neutral pions produce high-energy photons, detectable by HEX-P. These photo-pion processes are expected to happen in AGN jets \citep{Biermann1987,Mannheim1992,Mannheim_1993} and AGN cores \citep{Eichler1979,Berezinskii1981, Protheroe1983,Begelman1990,Stecker1991}. The photo-hadronic interactions likely responsible for the neutrino production in blazar jets inevitably lead to the development of electromagnetic cascades. These are initiated by synchrotron or inverse-Compton scattering by secondary  electron-positron pairs produced in photo-pion and Bethe-Heitler pair production processes, and the subsequent decay of charged and neutral pions and muons. 
The radiative output of these cascades is expected to peak in the hard X-ray -- soft $\gamma$-ray band \citep[e.g.,][]{2015MNRAS.447...36P,2018ApJ...864...84K,Murase2018,2019NatAs...3...88G,2021Physi...3.1098Z}.

Follow-up of IceCube neutrino alerts has been promising for identifying the source of astrophysical neutrinos and therefore cosmic rays.
The associations of blazars with VHE neutrino events detected by IceCube have attracted much attention in the recent literature, with increasing evidence pointing towards blazars being responsible for at least part of the IceCube VHE neutrino flux \citep[e.g.,][]{Krauss2014,Kadler2016,Garrappa19,Buson22,Plavin20,Plavin21,Plavin23}.
In 2017, a high-energy IceCube event (IceCube-170922A) was found in
positional and temporal coincidence with a blazar flare from TXS\,0506+056 \citep{txs}, 
following the previous, lower-confidence coincidence of a PeV neutrino 
event and a blazar flare \citep{Kadler2016}.  
Surprisingly, an earlier neutrino flare consistent with the position
of TXS\,0506+056 between September 2014 and March 2015 is not
accompanied by a $\gamma$-ray flare. In TXS\,0506+056, X-ray observations in the soft and hard X-ray band were crucial in constraining the hadronic contribution in order to determine that the source is a likely
neutrino counterpart.
These cascading signatures were studied in detail for TXS 0506+056 by \cite{Reimer19}, who used the measured VHE neutrino flux from the neutrino flare coincident with the direction of TXS~0506+056 in 2014 -- 2015 \citep{IceCube2018} to constrain the source parameters of the photo-hadronic interactions leading to the neutrino production. A crucial parameter in these calculations is the maximum optical depth to $\gamma\gamma$ absorption within the source (typically peaking at TeV -- PeV $\gamma$-ray energies), which determines the intensity of the cascades. 
In this case, the hadronic contributions to the electromagnetic SED were found to be small, but consistent with the neutrino
detection. Such small hadronic contributions are generally expected \citep{2018A&A...620A.174K}, as electrons and
positrons are much lighter, they are easier to accelerate and we
therefore expect a high fraction of leptonic emission from AGN jets.

Previous detections of possible blazar counterparts were followed by the detection of possible Seyfert AGN counterparts, that can only be identified in the X-ray band: IceCube-190331A \citep{2020MNRAS.497.2553K} and the recently discovered neutrino source NGC\,1068 \citep{2022Sci...378..538I}.
X-ray follow-up of likely-cosmic neutrino alerts has thus proven its utility for identifying possible neutrino counterparts, and most importantly for measuring hadronic
contributions to the high-energy emission.
Blazars have been ruled out as the dominant source of IceCube neutrino events in the TeV energy band through stacking analyses \citep[e.g.,][]{stacking3}. Blazars, especially BL Lacs are expected to produce neutrinos in the energy range of PeV to EeV \citep[e.g.][]{2014PhRvD..90b3007M, 2015MNRAS.452.1877P}. These are energies that will be covered by the Askaryan Radio Array \citep[ARA;][]{ARA}, currently under construction, as well as the IceCube-Gen 2 detector \citep{ICgen2}. IceCube Gen 2 with a 8 cubic kilometer volume is expected to be completed in 2032, which would coincide with the expected start of the HEX-P mission.

HEX-P will identify possible neutrino emitters and constrain the fluxes and emission mechanisms. One of the main challenges of neutrino follow-up is the large uncertainty regions of both cascade and track events. Track events are more promising for a direct identification of a counterpart source due to their smaller angular uncertainties, $\gtrsim 0.5^\circ$. This currently requires Swift-XRT tiling observations with at least 7 pointings to fully cover the uncertainty area. This can often be reduced by using knowledge of previously catalogued X-ray sources. eROSITA will be crucial in providing a pre-identification of potential counterparts. HEX-P has a field of view of $\approx 13^\prime \times 13^\prime$.  This will allow follow-up of known brighter X-ray  sources in only a few pointings.

\begin{figure}
\includegraphics[width=0.4\textwidth]{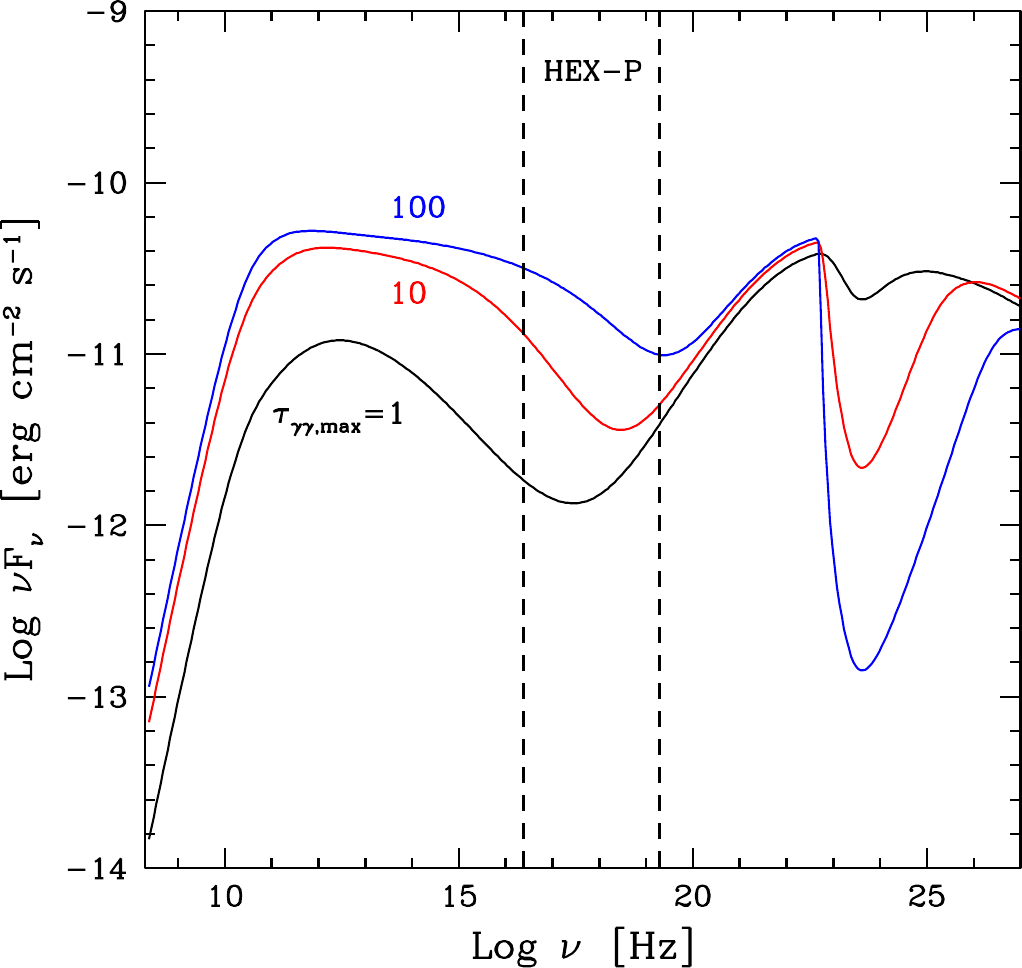}
\quad
\includegraphics[width=0.57\textwidth]{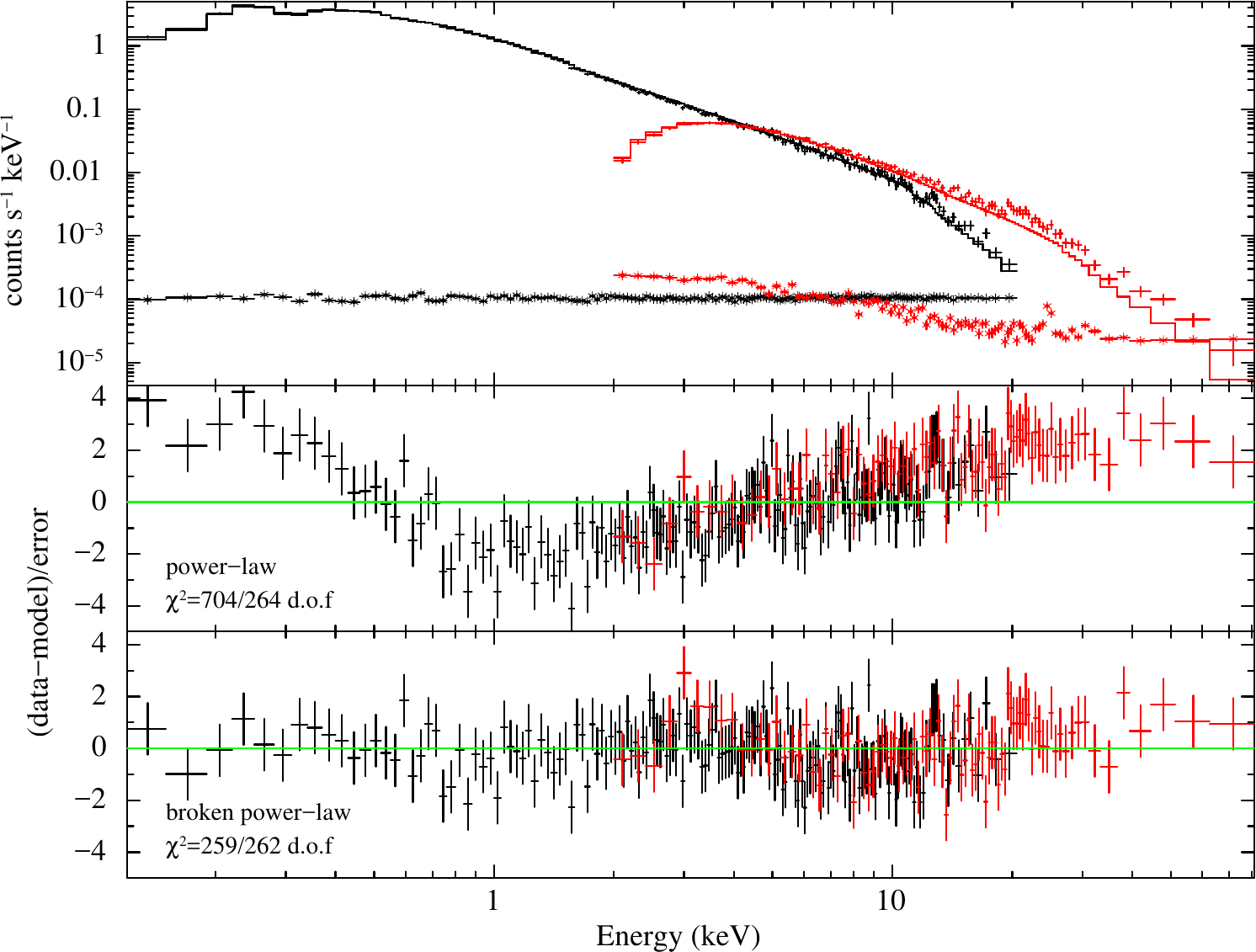}
\caption{\textit{Left panel:} Theoretical SEDs of TXS~0506+056 (including EBL absorption) calculated from simulations of photo-pion induced synchrotron-supported cascades, for values of $\tau_{\gamma\gamma}^{\rm max} = 1, 10$ and $100$; based on Figure 12 of \cite{Reimer19}.
Spectral breaks are present in the HEX-P band.
\textit{Right panel:} Simulated 50-ks observation of TXS~0506+056 for the model spectrum with $\tau_{\gamma\gamma}^{\rm max} = 10$ (red line in left panel). 
LET data are shown in black, HET in red. The background contribution is also shown in the upper plot.
The spectrum is fitted with pure power-law and  broken power-law models. The residuals indicate that the broken power-law is clearly preferred.  }   
\label{fig:Cascades}
\end{figure}

In order to test whether HEX-P observations are able to distinguish between a single power-law
and the curved up-turns expected from proton-induced cascades, we performed detailed simulations. Three model cascade spectra (synchrotron-supported cascades) from Figure 12 of \cite{Reimer19} (for $\tau_{\gamma\gamma}^{\rm max} = 1, 10,$ and $100$) were used as input to simulate 50 ks observations with HEX-P, which were subsequently fitted with both a single power-law and a broken power-law model. An example for $\tau_{\gamma\gamma}^{\rm max} = 10$ is shown in Figure~\ref{fig:Cascades}. In all three cases, a single power-law could clearly be rejected, while a broken power-law was strongly preferred.
Although this is not the only possible interpretation for such upturns, HEX-P can reveal and constrain this expected feature 
of the proton-induced cascades associated with photo-pion neutrino production.

\subsection{Variability Studies of Blazars }\label{sec:variability}
Variability is a defining feature of all blazar subclasses, and studies have been conducted across the entire electromagnetic spectrum for many years, both with a focus on long-term (weeks to years) as well as short-term (days to hours) time scales. It is increasingly recognized that simultaneous broadband variability study is crucial to understand blazar physics. But the sample sizes of these studies depend\st{s} largely on monitoring capabilities in the different energy bands.
In the soft X-rays,  monitoring campaigns can be performed for certain objects with Swift/XRT \citep{2019A&A...631A.116G,2021MNRAS.507.5690G}, while observations with, e.g., XMM-Newton enable investigation of variability on time scales of hours \citep[e.g.,][]{2006ApJ...651..782Z,2022arXiv220602159P,2022ApJ...939...80D}.
Long-term monitoring in the hard X-ray range is possible with Swift/BAT and MAXI (and in the past with RXTE), though, only for the brightest blazars. 
Variability studies above 10\,keV on intraday time scales have been performed with RXTE \citep{2018ApJ...867...68W} and later NuSTAR  \citep{2018A&A...619A..93B}. 
Since no existing X-ray telescopes have sensitive broadband capabilities, simultaneous coverage of soft to hard X-ray variability has to be performed by carefully coordinating multiple X-ray telescopes pointing, which has only been arranged for a small number of observations.

The combination of high sensitivity and time resolution on the order of milliseconds and up to 80\,keV will enable HEX-P to probe rapid variability, i.e., fluctuations within several minutes, at hard X-ray energies for a sample of bright blazars. As such time scales have not been probed before at $>10$\,keV, HEX-P observations will facilitate the discovery of new and potentially unexpected phenomena in blazar jets.
For example, in HSPs,  minute timescale variability  as seen is gamma-rays
which could become evident only in hard-X-rays, since produced by higher-energy electrons,
or a hardening at high energies as the result of the harder emission of freshly accelerated electrons on top of a more persistent, softer component.

In the case of FSRQs, HEX-P follow-up observations of flaring blazars will yield an unprecedented view up to 80\,keV, i.e., full coverage of the rising half of the high-energy hump, 
and enable time-resolved spectral studies, which might be able to track changes in the high-energy emission related to, e.g., a change from SSC to EC emission over the course of a flaring event. This part of the spectrum can be a mixture of SSC and EC under the leptonic scenario, while under the hadronic scenario it may consist of SSC, hadronic cascades, EC, or proton synchrotron \citep{Boettcher2013,Cerruti2015,Zhang2016}. Theoretical studies have shown that these radiation processes can have different variability patterns, which can lead to distinguishable spectral and temporal signatures in the broadband X-rays. For instance, in the leptonic model EC oftens peaks at higher energy than SSC, thus the X-ray bands consists of SSC by higher-energy electrons (thus more variable) and EC by lower-energy electrons \citep[see e.g.][]{ghisellini1999,Diltz2015,Zhang2015}. Simultaneous broadband X-ray variability can therefore track the transition from SSC to EC, as well as unveil possible 
bulk-compton emission components, which are expected to be transient \citep{celotti2007}.
They can thus probe the underlying particle evolution, and put constraints on the origin  of SSC and EC emissions.

\begin{figure}
\centering
\includegraphics[height=8cm]{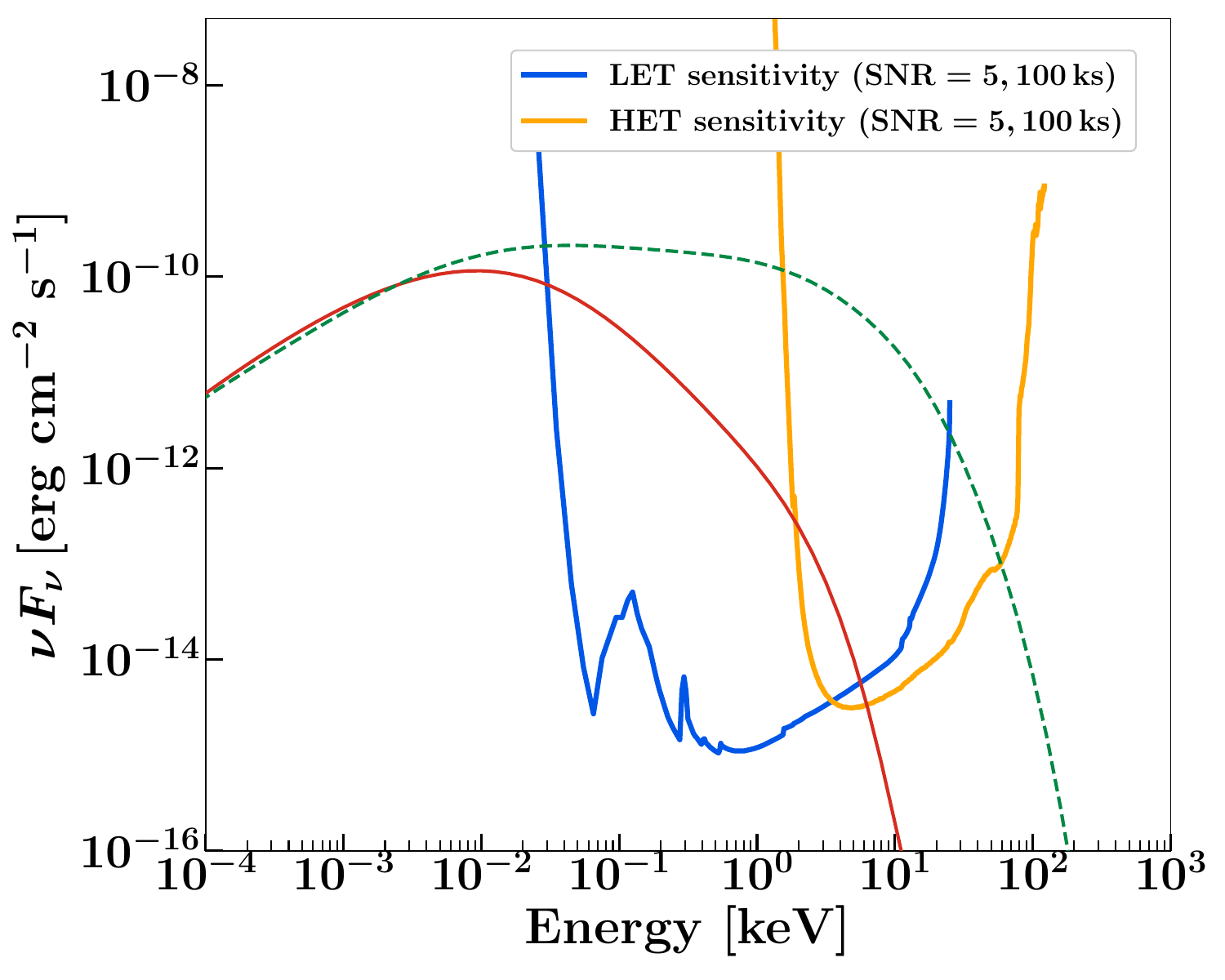}
\caption 
{ \label{fig:peak_shift}
Simulated synchrotron peak of a high-peaked BL Lac object in quiet state (red line) versus the shifted synchrotron peak from the same source during flaring activity (dashed green line). The sensitivity curves for both the LET and HET of HEX-P illustrate the excellent capability of the X-ray telescope to precisely measure changes in the position and shape of the synchrotron peak of blazars.} 
\end{figure}

Observations have shown that blazars of all types (low, intermediate or high-synchrotron-peaked objects) may experience epochs of extreme particle acceleration, resulting in its synchrotron peak shifting into or across the X-ray band.
Most dramatic events have been shown for example by Mkn 501 in 1997 \citep{pian98, tavecchio2001}, 1ES\,1959+650 in 2002 \citep{kraw1959} and more recently,  again by Mkn 501 \citep{veritas2016_501,magic2020_501}, Mkn 421 \citep{tramacere2009,magic2020_501}, PKS\,2005-489 \citep{tagliaferri2001,Chase2023} and 1ES\,1215+303 \citep{Valverde2020}. 
Such phenomena are often referred to as the peak shift. 
The physical cause is generally understood as the emerging of a 
new hard emission component from freshly accelerated electrons,
injected by a new dissipation event.
The mechanism and location of this acceleration is instead still uncertain,
being debated among shock acceleration (either in internal shocks or a standing shock),
magnetic reconnection and/or hadronic origins.
\citep{Boettcher2013,Zhang2022}. Due to the significant spectral changes, observing these events needs broadband coverage. 
Temporal slicing of the SED evolution is therefore important to elucidate the formation of such extreme physical conditions.
With the very good broadband sensitivity and timing capability (Figure~\ref{fig:peak_shift}), HEX-P can track the peak shifts and spectral variations in greater detail than previously possible, revealing the physical driver of such extreme particle acceleration and probe potential hadronic origins.
As such, HEX-P will serve as a space-based time domain astronomy instrument; a field that has been identified as one of the key research topics in the Astro2020 decadal survey\footnote{\url{https://cor.gsfc.nasa.gov/Astro2020/26141.pdf}} \citep{Astro_2020}.

For high-peaked BL Lac objects, spectral changes have been observed during flares, during which the spectrum often becomes harder when the source gets brighter. 
The study of light curves covering different energy ranges can reveal the existence of lags, which are indicators of the efficiency of the underlying acceleration and cooling processes.
Interestingly, these lags are not consistent: for example, the well-studied source Mrk\,421 has shown both soft and hard lags at different times, without, so far, a consistent pattern \citep{2010SCPMA..53S.224Z,Kapanadze_2016,2022arXiv220602159P}.
HEX-P will be an essential instrument for these studies, in particular for flare follow-up observations, as it covers both the soft and hard X-ray range with high sensitivity.

One key discovery in the last decades has been 
flux variations close to or below  the light-crossing time scale of the super-massive black hole  \citep[SMBH,][]{fast2155,fastIC310,fast3C279}, which revealed extreme particle acceleration in very localized regions. These rapid flux variations provide invaluable information about the particle acceleration mechanisms in jets and are broadly connected to the jet capability in accelerating protons and produce neutrinos.
Observed so far in a few exceptional flares, such rapid gamma-ray variability
has shown the limitations of standard emission models \citep{begelman08,ghisellini08,2019A&A...627A.159H}  and prompted more sophisticated scenarios \citep[see, e.g.,][]{aharonian17},
such as turbulent, multi-zone emission \citep{marscher14}, magnetic reconnection
inside the jet or in the jet launching region \citep[see, e.g.,][]{cerutti12,giannios13,kadowaki15,christie2019,Zhang2021},
jet-star interactions \citep{barkov12}, or acceleration in the black-hole magnetosphere \citep{levinson11}.
These scenarios predict different signatures in both spectral and temporal behaviors, requiring broadband time-domain observations to further understand the jet physics.
At VHE, CTAO is expected to probe broadband (sub-TeV to $>100$\,TeV) minute-scale variability in several other sources with less exceptional flares, as well as on 
sub-minute timescales for events like the one detected from the blazar PKS\,2155-304 in 2006 \citep{fast2155,sol2013}.


\begin{figure}
\centering
\includegraphics[width=\textwidth]{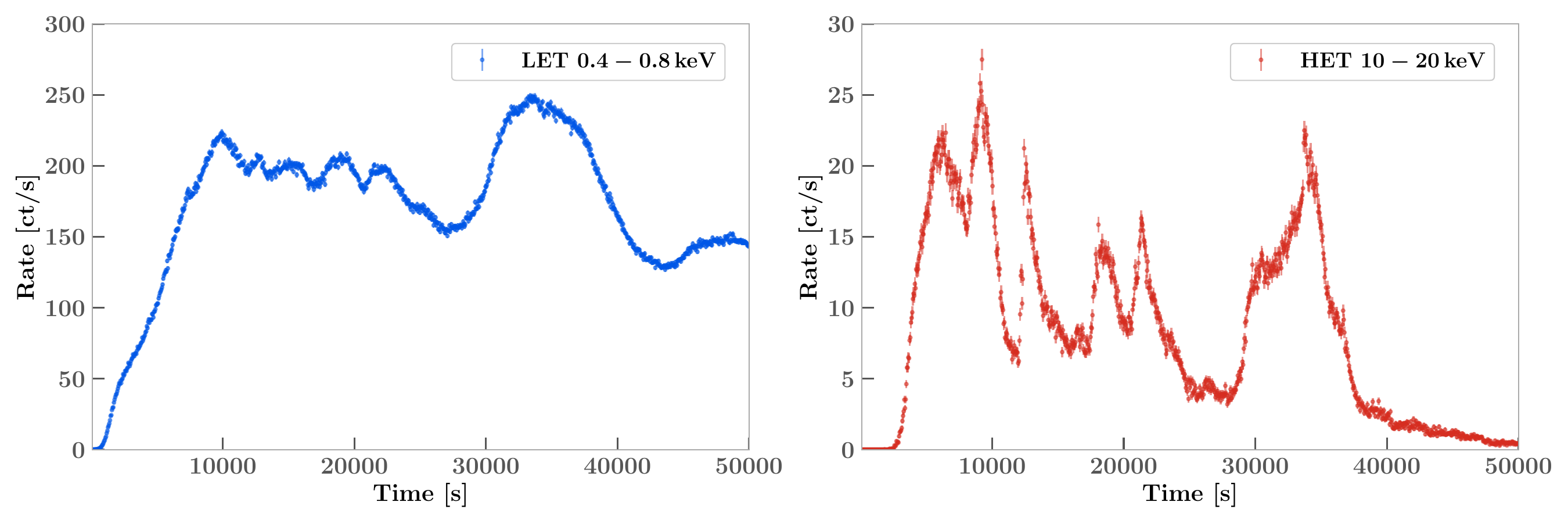}
\caption 
{ \label{fig:reconnection}
Simulated $50\,\rm ks$ light curves as detected by the LET ($0.4-0.8\,\rm keV$, left panel) and by the HET ($10-20\,\rm keV$, right panel) of a flaring blazar (such as Mrk 421) based on the magnetic reconnection model of \cite{Zhang2021}. The soft and hard X-ray bands show distinct flaring patterns, which can be detected by HEX-P to identify reconnection and constrain its physical conditions.} 
\end{figure}

Current theories suggest that such short timescale variability should also be present in X-ray synchrotron emission,
but so far the few simultaneous data show that, even when tightly correlated, 
the (soft) X-ray variations have  typically been of much lower amplitude than at VHE \citep{hess2155_chandra} and in general on longer timescales (tens of minutes).
While X-ray synchrotron emission may suffer from a stronger dilution of variable flux by brighter persistent emission,
a key issue in observation is that current X-ray instruments lack the same broadband sensitivity as the VHE telescopes. The synchrotron emission of the rapidly-variable TeV electrons could
peak at higher X-ray energies, in a band not yet investigated because of the lack of large collection areas above 5-8 keV.
Additionally, several models, such as magnetic reconnection, predict distinct variability patterns in soft and hard X-ray bands (Figure \ref{fig:reconnection}), which require broadband coverage with good temporal sensitivity \citep{Zhang2021}.
HEX-P will be able to probe, for the first time, such short timescales simultaneously at both soft and hard X-ray energies.

Another aspect is the different inter-band variability among X-ray bands.
The interplay between  acceleration, cooling, and escape timescales of electrons at a shock front or dissipation region can produce different and characteristic patterns in the variation of spectral index vs flux. For example, such variations can have different ``hysteresis loops" \citep[e.g.,][]{kirk98,kirk99,Bottcher_2002,Abeysekara_2017,Wang_2018,bottcher19,Chandra_2021}.
The timescales of the flux rise and fall provide information on the acceleration mechanisms, if nearly instantaneous or progressive,
and these could be different between hard and soft X-rays.  
So far, such studies have been limited to a few objects observed during large flares over one decade in energy.
HEX-P will be able to investigate these relations over more than two decades in energy, and at lower fluxes and/or shorter timescales.

\subsection{Synergies with Other Telescopes}\label{sec:synergies}
\subsubsection{CTAO}\label{sec:CTA}
In HSP blazars, electrons accelerated to TeV energies emit by synchrotron typically in the X-ray band and by SSC in the VHE band.
The possibility to perform extended simultaneous X-ray--VHE observations therefore 
provides a powerful diagnostic tool to understand particle acceleration and emission, 
allowing the evolution of a single freshly-accelerated electron population to be sampled by two different emission mechanisms. This is especially important
during flares with hour and sub-hour variability.  
To perform such studies, an X-ray satellite must have a large visibility window at any given epoch for most of the sky and must be able to point in an antisolar direction.
A large visibility window increases the chance that a source can be observed when it is found flaring,  while the antisolar direction enables simultaneous observations at night with ground-based telescopes when the target culminates at the highest altitudes above the horizon,
and thus can be followed by Cherenkov telescopes continuously under the best conditions for at least several hours. 
Such studies are not possible with XMM-Newton due to its strong solar panel constraints \citep{costamante04}. Chandra showed the potential for such studies with the detection of large flares by PKS 2155-304 in 2006, albeit with a much lower collecting area than XMM-Newon \citep{hess2155_chandra}.
At an L1 orbit, HEX-P will fulfill all these operational requirements, allowing us to perform such studies for the first time with a sensitive, flagship-class X-ray telescope.

\subsubsection{IXPE \& COSI}\label{sec:ixpe}
High-energy polarization, a recently opened new window to the Universe, will play a very important role in future efforts to understand the matter content and high-energy emission processes in jets. In a leptohadronic scenario, emission at the trough can be dominated either by proton synchrotron or synchrotron from secondaries \citep{Zhang2019}, and hence it can be highly polarized in the X-ray and MeV bands. Recent IXPE observations have demonstrated that determining the shape of the X-ray spectrum is crucial to correctly recover the polarization parameters, regardless if X-rays are part of the synchrotron or high-energy spectrum \citep{Ehlert2022,DiGesu2022,Liodakis2022,Middei2023,Middei2023ApJ}. This is particularly important for sources like TXS~0506+056, i.e., ISP blazars, where the X-rays lie in the trough between the low- and high-energy humps of the SED \citep{Peirson2022,Peirson2023}. Additionally, IceCube observations of TXS~0506+056 suggest that the hadronic spectral component, potentially peaking in the MeV band, can be variable \citep{IceCube2018}. Therefore, both the spectral shape and variability of the hadronic contribution can have significant effects on the X-ray to MeV polarization degree and angle \citep{Zhang2016}, therefore requiring broadband coverage in X-ray and MeV bands. However, IXPE and future X-ray polarimeters, like the enhanced X-ray Timing and Polarimetry mission (eXTP), have limited energy ranges. For example, IXPE is sensitive in the 2-8~keV band. On the other hand, future MeV missions, such as the Compton Spectrometer and Imager (COSI, \citealp{COSI}), observe in the MeV band ($0.2-5\,\rm MeV$), which cannot fully capture the rising part of the hadronic component (in the X-ray bands) and its variability. HEX-P's broad energy range will allow for accurate modeling of the X-ray spectrum and transition region in a large number of blazars, paving the way for detecting polarization from the jet's high-energy component.   Coordinated observations with missions such as IXPE and COSI in the future will enable measurements that can pinpoint or rule out hadronic processes by comparing the predicted and observed polarization degrees.

\subsubsection{SPT (CMB)}\label{sec:spt}
As discussed in 
Section~\ref{sec:variability}, truly simultaneous, broad-band monitoring is key in unraveling blazar jets structure as well as the emission mechanisms and thus particle acceleration processes operating in blazars.  One important spectral band is the millimeter regime, where, for the high-luminosity (FSRQ-type) blazars, the low-energy part of the SED peaks:  blazars are generally strong mm sources and their mm emission can even be used to find counterparts of un-identified Fermi-LAT gamma-ray sources \citep[see e.g.,][]{Zhang_2022}.   This is the spectral regime where the blazar synchrotron emission is believed to become optically thin, and thus reveals the entire volume of the jet rather than its surface.  

Time-resolved blazar studies in the mm band were usually conducted by individual radio / mm telescopes, generally dedicated to a single object at a time, and clearly revealed the large amplitude variability on time scales of days.  Such studies also revealed a strong degree of polarization (in some cases, more than 20\%), which is key in determining the orientation and uniformity of the magnetic field.  Current and planned Cosmic Microwave Background (CMB) facilities will
provide such data as a part of the CMB measurements:  while the astrophysical sources are ``noise" for CMB maps, the data -- obtained ``for free" -- are very valuable for studies of such sources.  This holds particularly true for blazars, since the current and planned CMB facilities collect data by continually scanning relatively large regions of the sky, and subsequently combinining the data to form sky maps.  Such time-resolved data result, generally, in a relatively well-sampled time series for many sources.  Importantly, these CMB facilities, in addition to flux, also collect polarization information, thus providing time series in both intensity and polarization.  One recent ``proof of concept" example is observations of the blazar PKS 2326-502 with the South Pole Telescope \citep[SPT;][]{Hood_2023} which revealed correlated flaring in the mm
(at 150 GHz) and GeV-range gamma-ray bands, measured with the Fermi-LAT.   

The next generation CMB facility is known as ``CMB Stage 4" and is planned for deployment in the early 2030s. This matches well with the proposed timeline for HEX-P.  CMB Stage 4, described in \cite{Carlstrom_2019} and \cite{Abazajian_2022}\footnote{\url{https://cmb-s4.org/}} is significantly more sensitive than the current CMB facilities and, in its full form, will cover well over 50\% of the sky.  The facility will have detectors sensitive over the range of 30 GHz to 280 GHz.  While the details of the instrumental sensitivity are still under development, it is expected that in 24 hours worth of sky scanning, CMB Stage 4 should be able to measure a source flux of 30 milliJansky (5\,$\sigma$).  This complements the blazar studies undertaken with HEX-P {\sl without altering the strategy of the anticipated CMB studies}, and such synergistic studies (with various classes of astrophysical sources, not just blazars) are already part of the instrument design and the data management plan.  
	
\section{Source classes (MeV blazars, extreme BL Lacs, and Masquerading BL Lacs)}\label{sec:source_classes}

\subsection{The Accretion Flow to Jet Power Connection} \label{sec:accre_fl}
Flat Spectrum Radio Quasars are defined as blazars with strong broad emission lines. They are generally associated  with strong jet emission, peaking at low frequencies both in synchrotron and inverse Compton components. Similar to low-peaked BL Lacs, their high-energy component dominates the X-ray emission, particularly the rising section of the hump. Depending on redshift, different parts of this emission are mapped with specific telescopes: among the most widely used X-ray instruments, Swift/XRT, XMM-Newton and the Chandra X-ray Observatory map the softer X-ray emission ($< 10$~keV), while Swift/BAT reaches 500~keV, and INTEGRAL reaches 10~MeV \cite[see e.g.\ ][]{2019A&A...631A.116G,Marcotulli_2022}. NuSTAR provides a sort of ``link", by covering the energy range between 3 and 80~keV \citep{2019A&A...627A..72G}. 
The two instruments onboard HEX-P guarantee an energy range that can overlap with all X-ray telescopes currently available to the astronomical community. 
With a single observatory, the inverse Compton component of low-peaked blazars and FSRQs can be mapped throughout cosmic time, by focusing on their evolution and possible variability, spanning across the whole inverse Compton emission of this AGN class. On the soft X-ray side, LET will be able to shed light on the amount of possible absorption or instrinsic breaks at soft energies. The nature of such soft break is still under scrutiny, and the recent analysis performed on different samples of FSRQs do not allow to constrain whether they are due to free escess aborption or intrinsic jet features, such as peculiar energy distributions of the electron involved in jets high-energy emission \cite[see e.g.][for a thorough discussion]{gaur21}. The significant increase in sensitivity given by the LET and HET combination will likely make a difference in discriminating between the two possibilities.  
The HET instead will extend spectroscopic studies currently done by Swift/XRT, Chandra and XMM-Newton up to significantly higher energies, getting simultaneously close to the inverse Compton peak. 
The wide spectral coverage will allow for the first time to uniformly characterize this section of FSRQ SEDs, crucial for the study and interpretation of jet physics and emission. 

Even though FSRQs have been known and studied for more than 60 years, many open questions remain about their relativistic jets: how do they relate to the accretion flow, whether their launching system can be associated with it, and how the accretion-jet system evolves across cosmic time. 
Better insight into the electron energy distribution and a better mapping of the spectral evolution of the inverse Compton emission in time will already be significant advancements in relativistic jet physics. 
Nonetheless, HEX-P's wide energy coverage and deep sensitivity have the potential to make even bigger improvements to our knowledge. 

The connection between accretion flow and the relativistic jet in blazars is in fact a key ingredient to understanding the process behind jet launch and acceleration. 
Relativistic jets are generally associated with geometrically thick accretion flows when their launching mechanism is modeled: an accretion flow funnel close to the central black hole, along with a strong magnetic field, are effective ingredients to accelerate plasma along the polar direction \citep{2011MNRAS.418L..79T}. 
In contrast, a geometrically thin accretion disk has not historically been considered a solid basis for an effective jet collimation. 
FSRQs show that a relativistic jet can coexist with an optically thick, highly UV-luminous accretion flow, since they show a distinct and luminous optical-UV thermal continuum that cannot be justified with jet emission, with strong and broad emission lines, along with powerful relativistic jets. Such optical-UV features are most likely interpreted as coming from a geometrically thin, optically-thick accretion disk ionizing gas clouds in its vicinity \cite[e.g.][]{2017A&ARv..25....2P}. 
Moreover, the direct correlation between jet and accretion power in blazars suggests a significant role in the jet launching process by the accretion disk, or at least in providing power for the overall process \citep{2008MNRAS.386..989J,2014Natur.515..376G}. 
The coexistence of these two structures is thus under deep scrutiny: coordinated multiwavelength observations of high-energy jet emission and optical-UV radiation from the disk are crucial to test the actual relation between disk and outflows. 
In particular, analogous to the cyclic X-ray binaries evolution, FSRQs might show variations in geometry and emission physics of the central part of their accretion flows, corresponding to jet launching or rebrightening. 
In case of an accretion rate fluctuation, a thin accretion disk might thicken towards its inner radii, providing a temporary funnel, allowing for significant magnetic field advection and ultimately accelerating a new emitting region in the relativistic jet. 
Coordinated fluctuations would also be expected during flux variations, and they would be easily identified during coordinated optical and X-ray monitoring.
On the other hand, no thick component might be necessary: in this case, the launching mechanism might be different from what we currently expect, and this specific multiwavelength simultaneous or correlated variability would not be observed. 
Coordinated optical-UV and X-ray observations can thus provide unique insights into the most powerful engines in the Universe. 
HEX-P is crucial in this sense, being able to thoroughly map the jet emission profile and its evolution over a broad energy range.
Changes in jet power or acceleration would be easily detectable thanks to changes in the inverse Compton profile, well explored by the combined effort of LET and HET.

\begin{figure}
\begin{center}
\begin{tabular}{c}
\includegraphics[height=9cm]{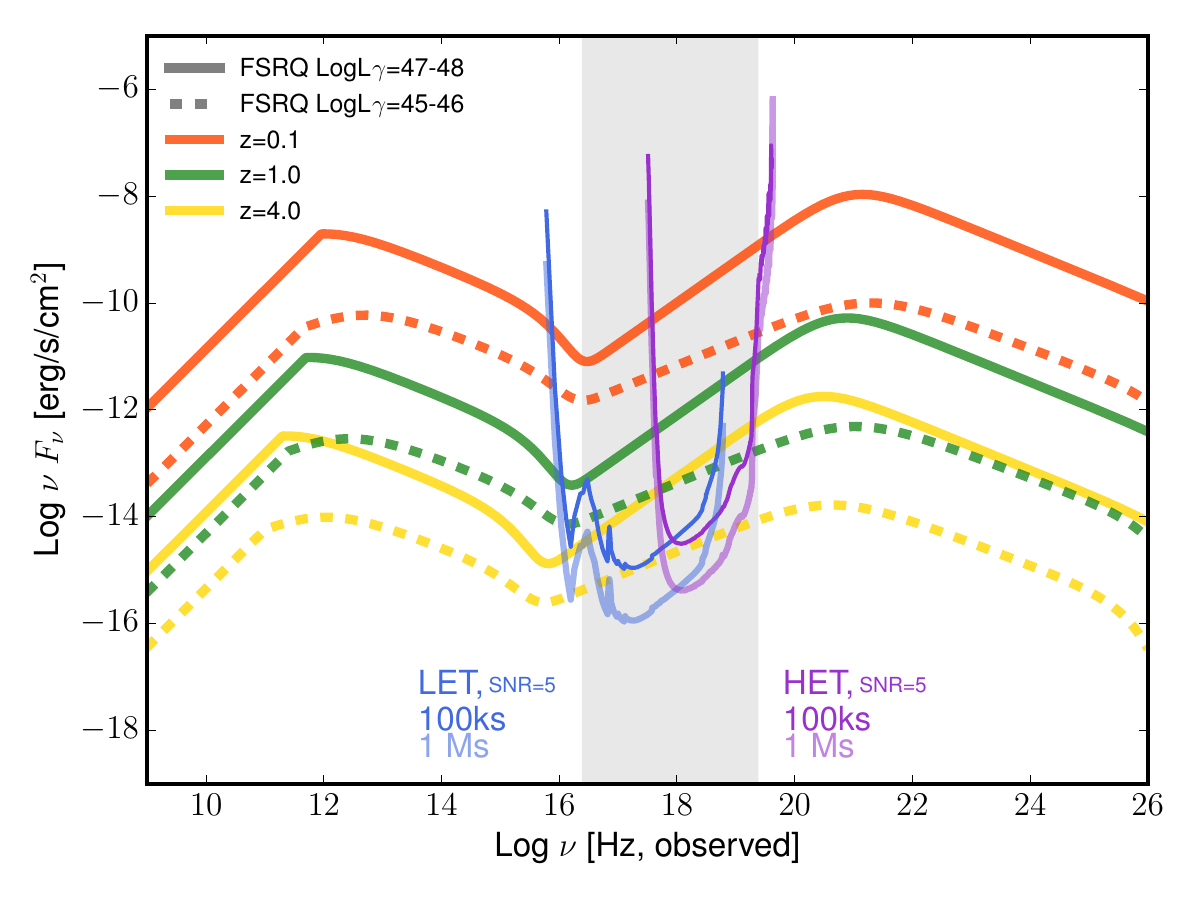}
\end{tabular}
\end{center}
\caption 
{ \label{fig:fsrq_sed}
FSRQ SEDs with different powers and redshifts, 
compared with LET (blue) and HET (purple) differential sensitivities, calculated with 100~ks (thin, bright) and 1~Ms (thick, faint) exposures.
The SEDs assume the Ghisellini et al.\ (2017) phenomenological models, corresponding to $L_\gamma=10^{47}-10^{48}\, {\rm erg}\, {\rm s}^{-1}$ (solid lines) and $L_\gamma=10^{45}-10^{46}\, {\rm erg}\, {\rm s}^{-1}$ (dashed lines). 
Different colours correspond to different redshifts. 
Note that the most powerful blazars will be easily detectable up to 80~keV and up to $z\sim4$. HET detections will be more challenging for less powerful FSRQs or sources in a low jet state. } 
\end{figure}

A good knowledge of the emission profile of high-energy blazar SEDs is crucial to nail the connection between jet and accretion flow in these sources. 
A reliable representation of the median blazar SED is necessary to identify variations in the jet emission, possibly linked to accretion fluctuations. 
The current state-of-the-art is mainly focused at low redshift.
Blazars with different jet powers are known to differ in their pre-peak inverse Compton slope and inverse Compton peak frequency. 
Even though the overall SED blazar shape is always characterized by the two prominent synchrotron and inverse Compton humps, their relative dominance and peak positions vary significantly depending on the jet power. Whether physically motivated or just an observing bias, blazars appear to peak at lower frequencies and have a harder X-ray spectrum when they are intrinsically more powerful.
This has been tested among the known $\gamma$-ray detected blazars, that are mainly concentrated below $z\sim3$, but a detailed study of the inverse Compton emission evolution across cosmic time is still incomplete. 
A significant limit is posed by the narrow X-ray coverage of the currently available sensitive X-ray observing facilities: hard X-rays are needed to explore the inverse Compton peak position, and no current telescope reaches deep enough to detect high redshift sources ($z>3$).
In this sense, HEX-P will be transformative.
For the very first time, the astronomical community will have the possibility to explore the evolution of AGN high-energy jet emission. 
At $z>3$, blazars appear to host among the most powerful jets known, with very high inverse Compton luminosities ($L_\gamma>10^{47}{\rm erg}\, {\rm s}^{-1}$) and MeV-peaking inverse Compton humps. 
This is likely a selection bias: we are currently  observing and classifying only the most powerful sources \citep{2020ApJ...889..164M,2021Galax...9...23S}. 
The possibility to consistently study most of the pre-peak inverse Compton SED over a wide redshift range ($1<z<4$), offered by HEX-P, will probe inverse Compton evolution across cosmic time. 
Limited (in numbers) but self-consistent (in method) results could already be drawn from a 1~Ms wide-field survey (see Section~\ref{sec:blazar_evolution}). 
Figure \ref{fig:fsrq_sed} shows how the expected HEX-P sensitivity limits and energy ranges will allow a comprehensive study of the most powerful FSRQs over a large redshift range.

\subsection{Supermassive Black Holes At The Dawn of Time}\label{sec:mev_bla}
Due to their high luminosities, we are able to trace blazars, particularly FSRQs, back to the early Universe. The entire broadband spectrum of these objects is shifted towards lower energies, and because of their high-energy hump peaking in the MeV band, they have been coined `MeV blazars' \citep{1995A&A...293L...1B}.
These sources are more luminous in the hard X-ray band \citep{2012MNRAS.426L..91S}, and studies made use of the NuSTAR satellite to analyse and characterise the emission processes responsible for the high-energy hump, in combination with Fermi-LAT at gamma-ray energies \citep{2013ApJ...777..147S, 2015ApJ...807..167T, 2017ApJ...839...96M, 2020ApJ...889..164M}.
The most puzzling findings about SMBHs hosted by quasars in the early Universe is that some of them exceed one billion solar masses, detected out to redshift of $z>7$ \citep{2001AJ....122.2833F, 2011Natur.474..616M, 2015Natur.518..512W}. Obviously, there is a bias towards the most luminous objects being detected, but the discovery of such massive black holes questions our understanding of how SMBHs form and how fast they can grow. Current theories involve either the creation of $\sim100\, M_\odot$ BH seeds as remnants of Population III stars \citep{2001ApJ...551L..27M, 2002ApJ...571...30S}, which then require sustained super-Eddington accretion growth periods \citep{2017MNRAS.464.1102B, 2020ApJ...904..200Y, 2022ApJ...934...58J}, or the direct collapse of gas into $10^4-10^6\, M_\odot$ SMBHs in pre-galactic discs or halos \citep[e.g.,][]{2003ApJ...596...34B, 2006MNRAS.370..289B, 2006MNRAS.371.1813L}. Another hypothesis is the formation of massive BH seeds within star clusters with masses of $\sim1000\,M_{\odot}$, which then grow due to the copious supply of matter in the star cluster \citep{2009ApJ...694..302D, 2014Sci...345.1330A}.

A fraction of these SMBHs will power jets that, if pointing towards us, should appear as MeV blazars. In particular, the detection of these sources in the gamma-ray band is limited by the sensitivity of current instruments, 
paired with a very low flux due to the immense distances. 
Less than 30 high-redshift ($z\geq2.5$) blazars have been detected at gamma-ray energies with \textit{Fermi}-LAT \citep{3FGL, Ackermann_2017, Liao_2018, 4FGL, Kreter_2020}, with the most distant source having a redshift of $z=4.72$.  
In addition, EBL absorption attenuates high-energy gamma-ray emission above $\sim$\,1\,GeV \citep{HESS_2013,Ackermann_2017}.
The large majority of high-$z$ blazar detections at gamma-ray energies were only possible due to flaring episodes \citep{Paliya_2019b, Kreter_2020}, which biases our studies towards more variable sources.
At X-ray energies, instruments like \textit{ROSAT} and \textit{Swift}/BAT have been able to perform unbiased all-sky surveys, but have a limited sensitivity. Pointing observatories like \textit{Chandra}, \textit{XMM-Newton}, and \textit{NuSTAR} have been used to study a selected sample of high-redshift blazars \citep[e.g.,][]{Bassett_2004, Young_2009, Ighina_2019}. In particular, \textit{NuSTAR}, with its sensitivity reaching up to 80\,keV, has significantly contributed to the study of high-$z$ blazars by providing hard X-ray data that is essential to constrain the shape and peak of the high-energy hump of the blazars' SEDs \citep[e.g.,][]{2017ApJ...839...96M, 2020ApJ...889..164M, Paliya_2020, Middei2022}. Similarly, HEX-P will enable us to continue studying blazars in the early Universe.


AGN create feedback mechanisms in their host galaxies and can be linked to their evolution \citep{1998A&A...331L...1S}. On large scales, jets influence their surroundings by creating shocks in the surrounding gas \citep{2000MNRAS.315..371S, 2004MNRAS.355L...9R, 2011MNRAS.413.2815S,2017MNRAS.471..658R}, leading to truncation of star formation. Studying AGN-galaxy co-evolution in the early Universe has been difficult and so far has only been done up to redshift $z=4$ for jetted AGN \citep{poitevineau2023}.
Due to its superior sensitivity compared to current X-ray telescopes, HEX-P will detect sources with fluxes as low as $10^{-16}$ erg cm$^{-2}$ s$^{-1}$.
Simulations in \citet{Civano23} show that the HET will be able to detect between 50 to 1000 sources (depending on the chosen evolutionary model, see \citealp{Marcotulli_2022}) 
in a 2\,Ms wide-field survey. This will include a large number of newly detected extragalactic objects, including high-redshift blazars. The discovery of new sources will be less biased towards the most luminous sources, and enable us to get a better understanding of the cosmic evolution of blazars (see Sect.~\ref{sec:blazar_evolution}), while hard X-ray data above 20\,keV will help distinguish the different high-energy processes in blazar jets, and allow comparisons of local and early Universe objects.

\subsection{Extreme BL Lacs}\label{sec:extreme_bllac}
Observations in the last two decades have shown that BL Lacs can reach 
much higher SED peak energies than previously thought,
surpassing by 1 to 3 orders of magnitude or more the values of the 
bulk of the HSP population.  
They have been thus called {\it extreme BL Lacs} \citep{costamante01}.
They come in two flavours: extreme synchrotron and extreme-TeV objects, depending on whether
they show extreme peak energies in the synchrotron or  gamma-ray emission.
By convention, an HSP blazar starts to be called extreme if the $\nu_{peak}$ is above 1 keV or 1 TeV, respectively, but these sources have displayed peak energies in excess of 100 keV 
\citep[both in flaring and quiescent states;][]{pian98,costamante01}  
and 10 TeV \citep{hess0229}, respectively.

Conventional one-zone leptonic models (see also Section~\ref{sec:lep_mod}) struggle to model SEDs peaking at such high-energies, especially in gamma-rays.
They generally require extreme conditions with very low magnetic fields,  several orders of magnitude below equipartition, 
unusually hard electron distributions and very low radiative efficiency 
\citep{costamante2018,biteau2020,zech2021}.
The relation between extreme synchrotron and extreme TeV emission is still not well established:
they should be correlated in a standard SSC scenario,
but there are objects which are extreme in synchrotron but not in gamma-rays, and vice-versa,
although extreme-TeV sources do tend to display also extreme synchrotron peaks more often than the other way round \citep{biteau2020}.
Some objects show large peak variability, both in amplitude as well as in energy,
with shifts by more than 3 orders of magnitude during flares 
\citep[see, e.g., Mkn 501,][]{magic501_2012,magic501_2014}.
Others instead show rather steady extreme features over several years, 
a fact that could hint at the contribution of different mechanisms to the gamma-ray emission, 
in addition to inverse Compton emission by ultrarelativistic electrons \citep{biteau2020}.

The location of the synchrotron peak in the SED and the slope of the spectrum around it are essential
parameters  to constrain the emission models and to study the origin of the peak itself
(if due to cooling, escape, or the end of the accelerated particles' distribution).
So far, this has been possible only for peaks up to $2-5$\,keV with great sensitivity, 
or up to $30-50$\,keV for only the brightest objects.

\begin{figure*}
    \centering
    \includegraphics[width=0.45\textwidth]{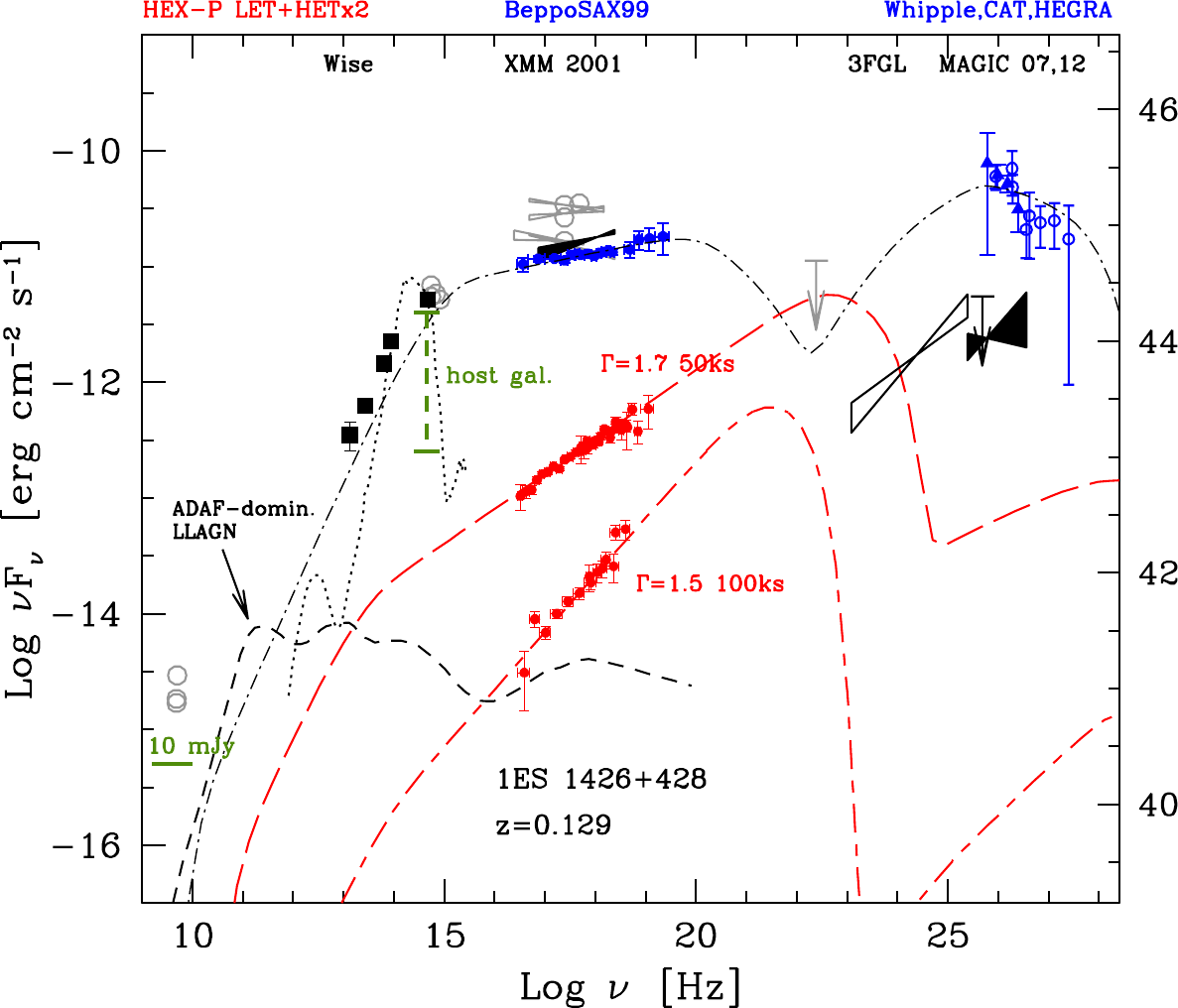}
    \includegraphics[width=0.45\textwidth]{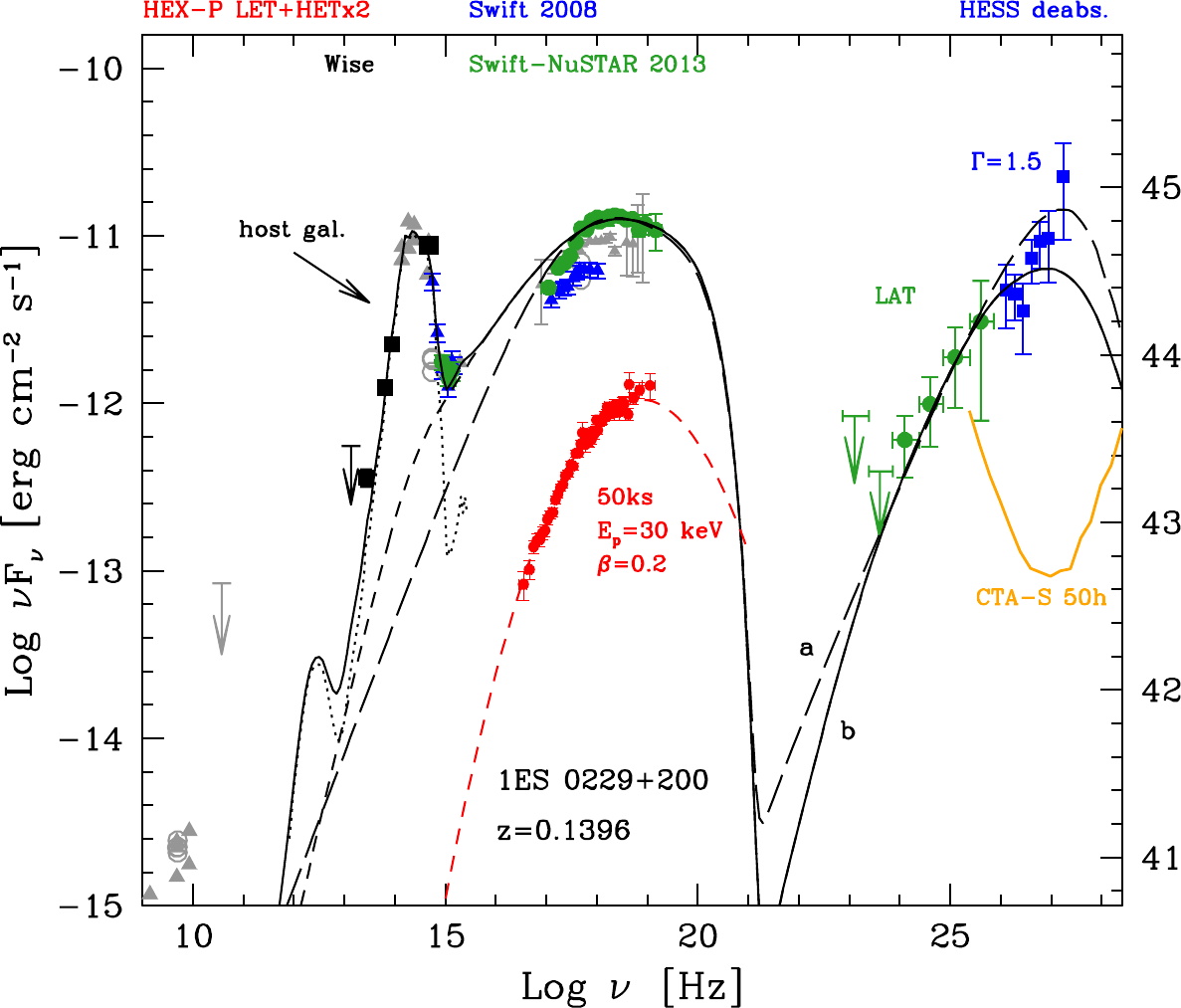}
    \caption{HEX-P simulated SEDs of two prototypical extreme BL Lacs, assuming lower flux levels. {\bf Left:} 50 ks and 100 ks simulations of two extreme sources with hard power-law X-ray spectra, possibly peaking in the MeV range \citep[adapted from][]{ghisellini99a, costamante07}. For reference, the figure shows also the average SED of an advection-dominated accretion flow low luminosity AGN \citep{nemmen14} at $30 \times$ higher luminosity, and in the optical the expected range
    of luminosities for the host elliptical galaxy (from 1ES1426+428 itself to a typical AGN-hosting galaxy).
    {\bf Right:} 50 ks observation of a 1ES\,0229+200-like object, but peaking at 30 keV with curvature $\beta$=0.2 and at a flux level  one order of magnitude fainter.}
    \label{fig:extremes}
\end{figure*}

Thanks to the unique combination of broad-band coverage, low background, and large collecting area from 0.2 to 80 keV
with a single observation, HEX-P is ideally suited to identify and study these objects:
it can directly measure or constrain the peak emission at flux levels or timescales 
one to two orders of magnitude lower than any current instrument (see Fig. \ref{fig:extremes}).

In this respect, the synergy with the upcoming CTAO---which will bring a similar improvement of one to two orders of magnitude in sensitivity at multi-TeV energies---cannot be understated.
Besides characterizing VHE spectra over a wide energy band, 
simultaneous observations  can test if and when these are indeed the same electrons emitting by two different mechanisms (synchrotron and SSC), as indicated by the tight correlations observed in HSPs down to minute timescales in Mkn~421 \citep{maraschi99,fossati2008,balakovic2016,magic421_2017} and PKS\,2155$-$304 \citep{hess2155_chandra}. CTAO and HEX-P coordinated observations will also test if the same relations that hold in extreme BL Lacs are observed in normal HSPs. 

A crucial  advancement enabled by HEX-P
is in population studies of extreme blazars.
Their overall luminosities tend to be low ($\sim10^{44-45}$ erg\, s$^{-1}$) compared to regular HSPs,
more or less in agreement with the blazar-sequence framework \citep{Ghisellini_1998,Ghisellini_2017}.  
So far, they have only been found at low redshift (up to $z\sim$0.3).
However, it is not clear if this is an intrinsic characteristic or simply a bias due to the low fluxes and limited observations.

With 50 ks exposures, HEX-P can discover and characterize such sources down to a $2-10$\,keV flux of $F_{2-10\,\mathrm{keV}}=10^{-13}$ \,erg\,cm$^{-2}$s$^{-1}$ out to $\sim$30\,keV. For example, with a photon index $\Gamma=1.8$, the 1-sigma 
error on the slope would be $\pm0.04$ full band, and $\pm0.20$ above 3\,keV (vs $\pm1.0$ with NuSTAR, for the same exposure time---and twice the clock time given the inefficiency of NuSTAR's low-Earth orbit compared to HEX-P's L1 orbit).
With $\Gamma=1.6$ and $F_{2-10\,\mathrm{keV}}=5 \times 10^{-14}$\,erg\,cm$^{-2}$s$^{-1}$, an extreme BL Lac can still be detected up to $\sim20$ keV with a slope uncertainty of $\pm0.08$ full band and $\pm0.25$ above 3 keV (compared to $\pm1.2$ with NuSTAR).
HEX-P can therefore discover and characterize sources similar to 1ES\,1426+428 (i.e., $L\sim10^{44}\,\rm erg~s^{-1}$) down to a luminosity of 
1$\times10^{43}$ erg/s, or up to $z=0.7$.

For objects with peak fluxes around $10^{-12}$\,erg\,cm$^{-2}$s$^{-1}$, HEX-P can also fully characterize the curvature of the synchrotron emission. For example, with a SED peaking at 30 keV 
and a curvature parameter $\beta=0.2$ (typical of HSPs), 
HEX-P can constrain the peak energy to better than 30\% and the curvature to within 10\% (Fig. \ref{fig:extremes}).

CTAO is expected to discover many more TeV-emitting HSPs, as allowed by a factor 10x improvement in sensitivity at 1 TeV with respect to present arrays.
Just from the planned CTAO extragalactic survey, which will cover $\sim$25\% of the sky,  about 110-180 new detections are expected \citep{CTA}, 
about 1/4 of which could be extreme-TeV according to the present number of extreme objects among TeV-detected HSPs \citep[see e.g.][]{costamante2020, biteau2020}.
Above a radio flux of 3.5 mJy at 1.4 GHz, the inferred surface density of 
$\sim4.5\times10^{-3}$ deg$^{-2}$  for extreme-synchrotron objects corresponds to a total number over the sky of $\sim$200 \citep{biteau2020}. 
Searches for these extreme blazars will benefit from the eROSITA all-sky survey. This survey is expected to have a point-source sensitivity of $6 \times 10^{-15}$\,erg\,cm\textsuperscript{-2}\,s\textsuperscript{-1} in the $0.2-2$\,keV band \citep{eFEDS}, making it more than ten times deeper than the ROSAT all-sky survey \citep{rosat_2rxs}. Additionally, eROSITA  will offer enhanced sensitivity in the $2-5$\,keV band, enabling the identification of further extreme blazar candidates. This will be particularly advantageous for low-luminosity extreme HSP BL Lac objects, which can then be confirmed and characterized by HEX-P.

\subsubsection{MeV-synchrotron BL Lacs}\label{sec:mev_synch_bll}
In principle, blazars could reach even higher energies than shown by extreme BL Lacs.
Expressing the acceleration time $t_{acc}$ in terms of the gyroradius $r_g$, 
$t_{acc}=\eta(E)\; r_g/c$ \citep[see e.g.][]{felixBook}, where the factor $\eta(E)\geq1$ 
characterizes the energy-dependent rate of acceleration
(where unity corresponds to the maximum possible rate in shock acceleration, 
when a particle gains all its energy in 1 gyroradius).
Therefore, even extreme BL Lacs (where $h\nu_{peak}/\delta\sim 1-10$ keV,
with $\delta$ being the beaming factor)  are in fact {\it not} extreme accelerators ($\eta\gtrsim 10^4-10^5$). 
In the Crab nebula, for example, unbeamed synchrotron emission up to $10-20$ MeV is observed, 
meaning electrons are accelerated at $\sim10\%$ of the maximum rate.

There might be blazars where
the acceleration  reaches even higher rates than observed so far, yielding a synchrotron peak in the MeV range \citep{ghisellini99a,ghisellini99b}. 
For example, $h\nu_{peak}\sim6-300$ MeV for $\eta\approx20-1000$ and $\delta=10-40$.
The luminosities are expected to be at the low end among blazars, if these objects follow the blazar sequence trend.
Such objects could have easily escaped detection so far.
At radio wavelengths, the flux is expected to be weak, likely below the typical level
of large-area sky surveys or the selection cuts for blazar samples.
In the optical band, the non-thermal continuum could be hidden one to two orders of magnitude below
the host-galaxy thermal emission. 
In soft X-rays, the jet emission is expected to be low,
thus possibly hidden under (or mixed with) the emission of the corona. 
In the VHE band, the SSC flux is likely too faint for detection, with the emission severely inhibited
by Klein-Nishina effects.
In summary, they could be faint at all frequencies except the hard X-ray and MeV bands, 
where they emit all their power and could reach luminosities $\sim10^{43-45}$ erg s$^{-1}$.
As a result, even if already detected in present or future deep X-ray surveys, such sources might not be recognized as blazars, having been more likely classified as radio-quiet AGN.

The brightest among such objects could be detectable by COSI, but only HEX-P can pin-point and reveal 
the exceptional jet emission and extreme accelerator inside otherwise unremarkable objects.
HEX-P can identify such a source with a peak luminosity of $10^{43}$ erg s$^{-1}$ up to $z=0.5$,
with the observational signature of a hard X-ray spectrum  ($\Gamma=$1.5-1.7),
connecting by power-law to the possible COSI (or even Fermi-LAT) detections.

\subsection{Masquerading BL Lacs}\label{sec:masquerading}
Masquerading BL Lacs are objects that are classified as BL Lacs because of the lack of broad emission lines in their optical spectra, but are very likely FSRQs in disguise \citep{giommi2013}. In these objects the jet is well aligned with our line of sight and its synchrotron emission overwhelms the disk and broad line region (BLR) emission, making the latter undetectable. The community has shown renewed interest in them after TXS~0506+056, a BL Lac associated with the high-energy neutrino IceCube-170922A, was revealed to be a masquerading BL Lac \citep{padovani2019}. Indeed, neutrino emission in blazars via photo-hadronic interactions requires external photon fields, something that a masquerading BL Lac can provide. In the last few years, more masqueranding BL Lacs potentially associated with IceCube neutrinos have been discovered \citep{padovani2022a,padovani2022b}.
Distinguishing a BL Lac from a masquerading one is often complex and requires SED modeling with high quality data. This becomes somewhat easier when sources are detected at high redshift ($z>1$). In that case, the large radio power and Compton dominance gives away their true nature \citep{rau2012}.  High-redshift masquerading BL Lacs have also been dubbed `blue FSRQs' \citep{ghisellini2012} because of their blue SED which typically shows synchrotron peak frequencies greater than $10^{15}$\,Hz and very hard $\gamma$-ray spectra. This implies that the electrons experience weak radiative cooling, a condition satisfied if the location of the emission region is outside the BLR \citep{ghisellini2012}. The hard $\gamma$-ray spectra and large distances make masquerading BL Lacs powerful probes of the extragalactic background light \citep[][]{ebl2012,ebl2018}, the integrated emission of all star-forming activity in the Universe.

\begin{figure}
    \centering
    \includegraphics[width=1\textwidth]{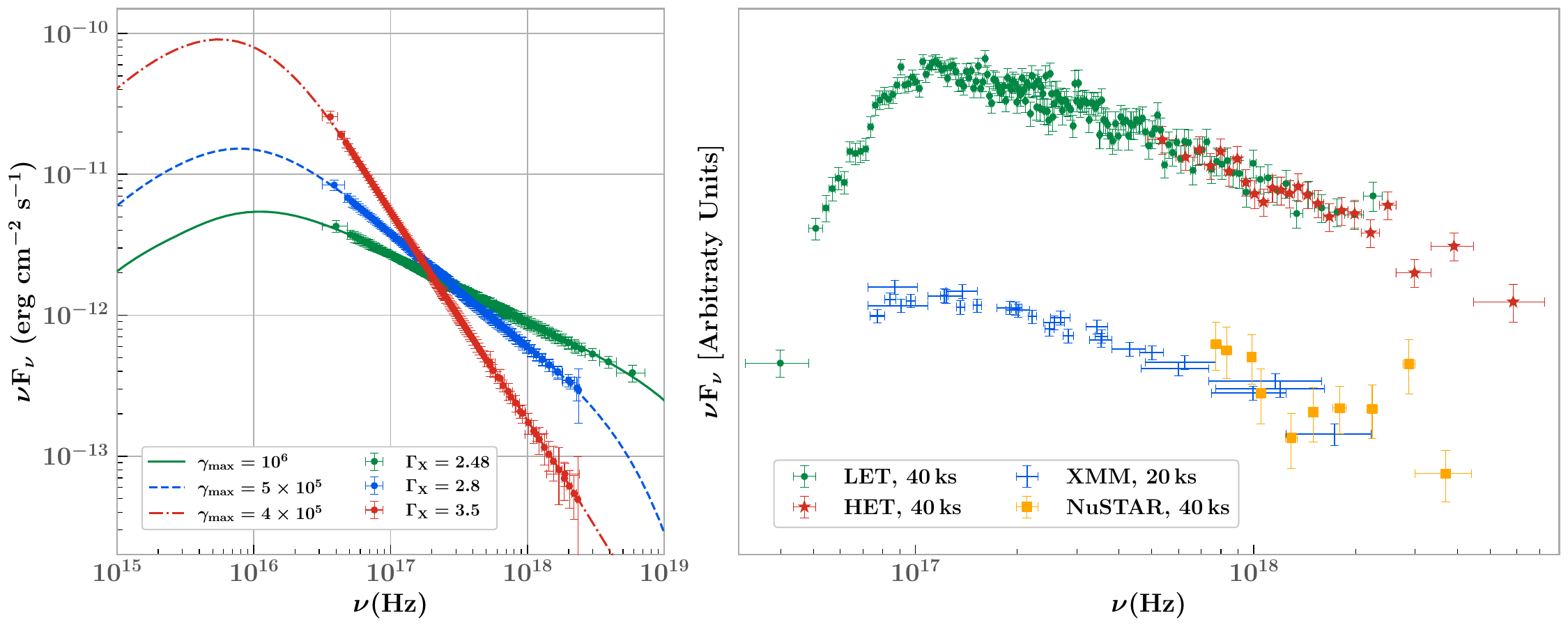}
    \caption{{\bf Left:}  Simulated SED of the high-redshift BL Lac 4FGL J2146.5-1344 ($z = 1.34$; \citealp{Rajagopal_2020}). The left panel shows the simulated HEX-P SED with different spectral indices ($\Gamma_{\rm X}$). In only $40\rm\,ks$, HEX-P will characterize the source spectrum from $0.2\,\rm keV$ to $30\,\rm keV$. This will provide crucial constraints on the shape of the high-energy particle population (e.g., the maximum Lorentz factor, $\gamma_{\rm max}$), the total jet power, and the true nature of the emitter (whether it is a high-luminosity BL Lac or a masquerading FSRQ). {\bf Right:} comparison of $40\rm\,ks$ HEX-P simulations to observations carried out by \citet{Rajagopal_2020}, which entailed simultaneous observations of 4FGL J2146.5-1344 with XMM-Newton ($20\,\rm ks$) and NuSTAR ($40\,\rm ks$). HEX-P data quality for a similar exposure will highly enhance our S/N and improve determination of important spectral parameters (such as  photon index and flux) as well as enable us to detect spectral curvature beyond $10\,\rm keV$, which would require a $> 100\,\rm ks$ NuSTAR exposure (and $>200\,\rm ks$ clock-time, given NuSTAR's low-Earth orbit). }
    \label{fig:masq_bll}
\end{figure}

HEX-P's deep, broadband sensitivity is a unique advantage compared to other instruments for studying masquerading BL Lacs. By measuring the synchrotron component to high energies, HEX-P 
can put constraints to the maximum Lorentz factor ($\gamma_{\rm max}$) of the electron population, at least up to the energy where the inverse Compton emission starts to dominate. This is something that is generally unconstrained given the limited sensitivity of X-ray instruments above $40$\,keV. In Figure~\ref{fig:masq_bll} we show an example of a simulated SED for the high-redshift ($z=1.34$) BL Lac 4FGL J2146.5-1344 studied by \citet{Rajagopal_2020}. In just $40\,\rm ks$, HEX-P will be able to characterize the X-ray range in the SED of the source from $0.2\,\rm keV$ up to $30\,\rm keV$, enabling us to accurately determine parameters such as the electrons' $\gamma_{\rm max}$ and to detect a possible break in the high-energy spectrum of the source. We note that to achieve the same quality spectrum $>10\,\rm keV$, it would require $>100\,\rm ks$  NuSTAR observations (and $>200\,\rm ks$ clock-time, given NuSTAR's low-Earth orbit). 
As extensive effort from the community has been dedicated to measuring the redshift of BL Lac sources via photometric and/or spectroscopic follow-ups  \citep[e.g.][]{shaw13,massaro14,paggi14,ricci15,landoni15,marchesini16,alvarez16a,alvarez16b,Massaro_2016,marchesi18,deMenezes_2020,desai2019,Pena-Herazo_2019, Pena-Herazo_2021, rajagopal2021, Sheng_2022}, the sample of high-$z$ BL Lacs is bound to increase in the next decade. These will be prime targets for a mission like HEX-P.

\begin{figure}
    \centering
    \includegraphics[width=0.47\textwidth]{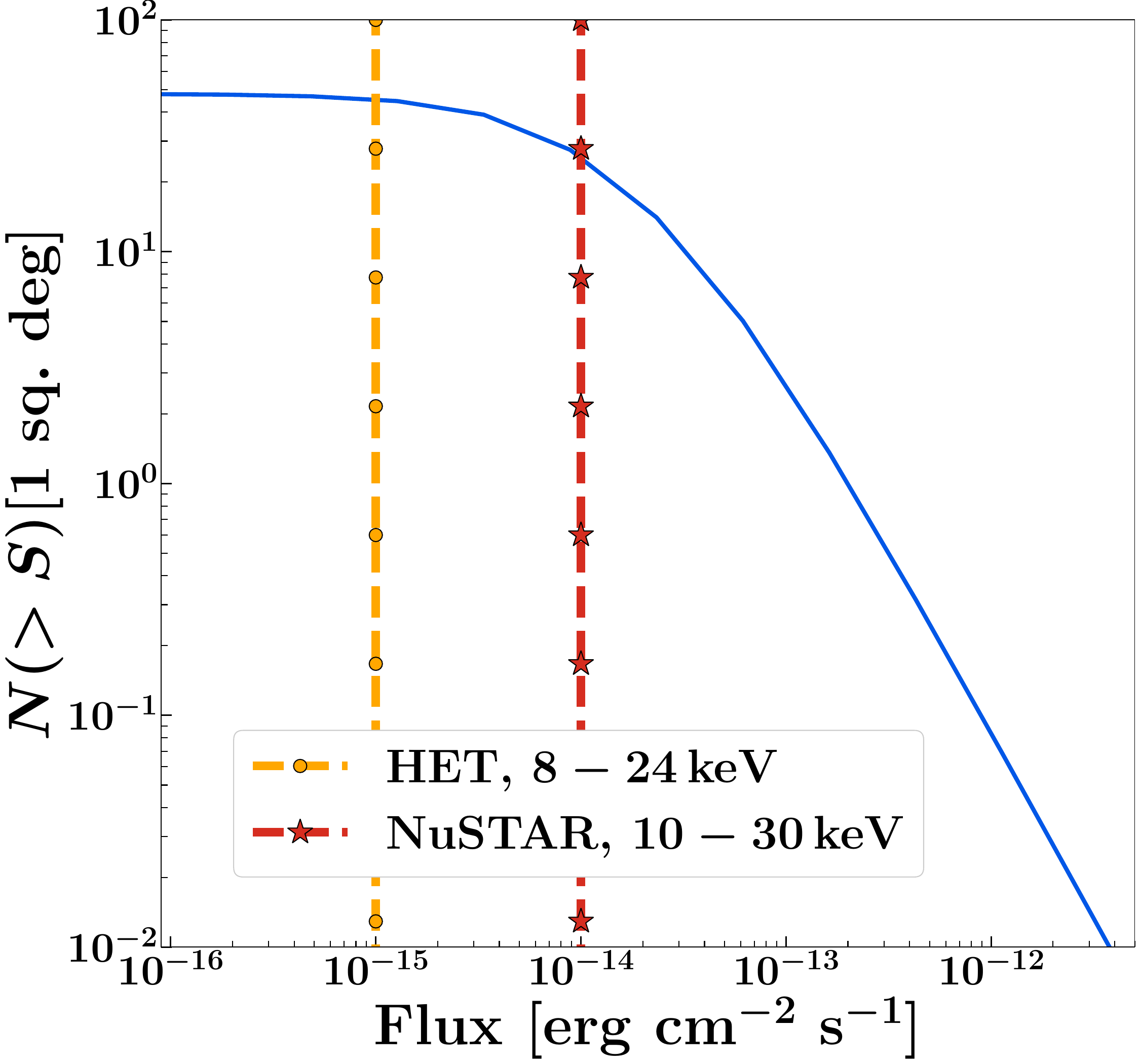}
    \includegraphics[width=0.47\textwidth]{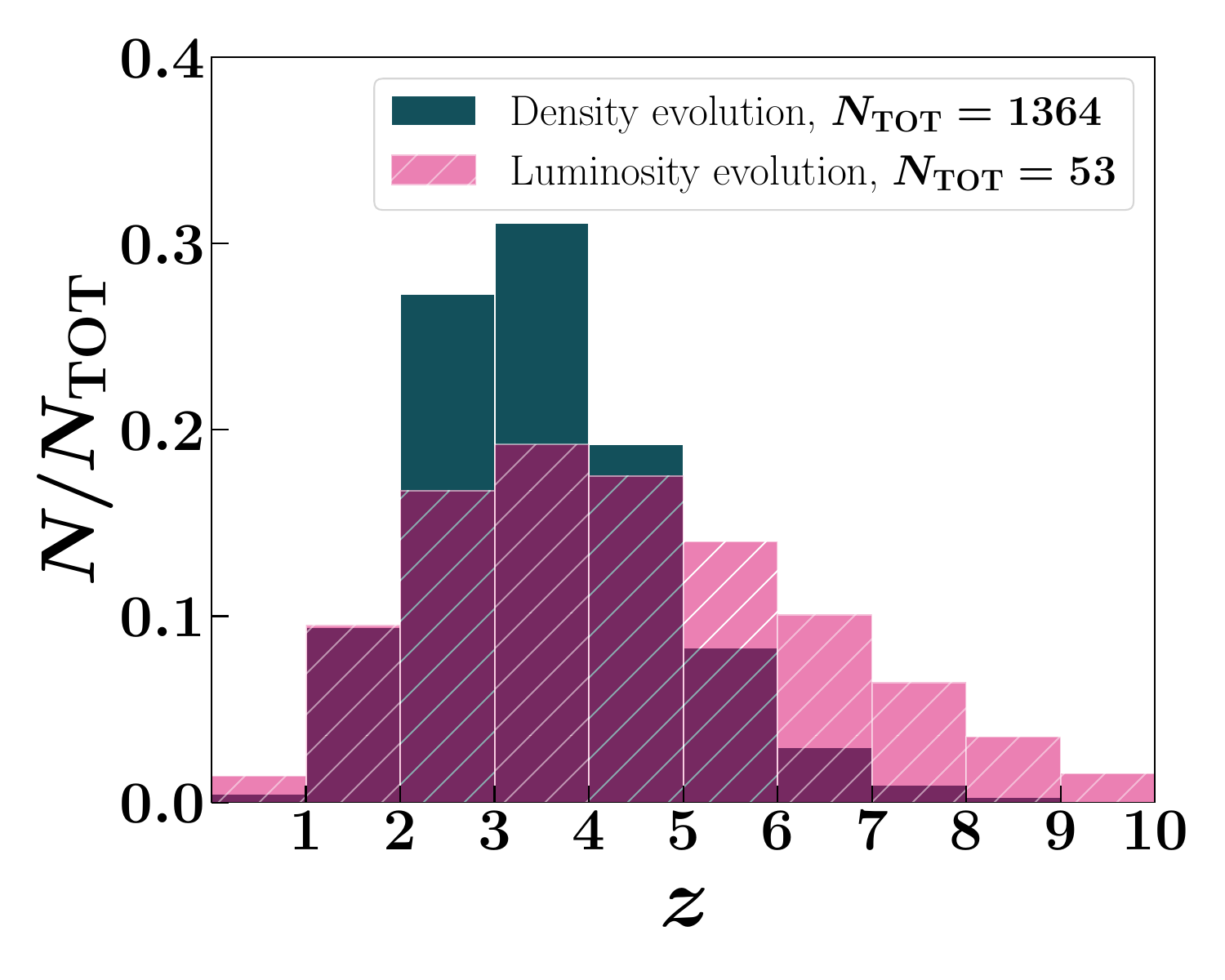}
    \caption{{\bf Left:} Blazar source count distribution, {\it logN-logS} (blue line), derived from the best-fit luminosity evolution model of \citet{Marcotulli_2022} expected for a 1 deg$^2$ survey. The yellow dash-circle line indicates the HET $8-24\,\rm keV$ source sensitivity for the planned 1 deg$^2$ wide survey \citep{Civano23}. For comparison, the red dash-star line represents the NuSTAR source sensitivity in the $10-30\,\rm keV$ band. HEX-P sensitivity would allow us for the very first time to detect the break of the luminosity function (which is yet to be seen by current facilities). This would lead us to definitely constrain which kind of evolution X-ray detected FSRQs have undergone, as measuring such break would solve the degeneracy between a density versus a luminosity evolution. {\bf Right:} Predicted relative number of FSRQs ($N/N_{\rm TOT}$) that would be detected by HEX-P in the planned 1 deg$^2$ wide survey as a function of redshift. Prediction using the best-fit density evolution model are shown as solid dark green bars, and using the best-fit luminosity evolution model as hatched pink bars \citep{Marcotulli_2022}. Depending on the number of sources detected by HEX-P as a function of redshift, we would be able to determine for the very first time the evolution properties of the population.}
    \label{fig:logn_logs}
\end{figure}

\subsection{Blazar Evolution}\label{sec:blazar_evolution}

As a source class, blazars have been shown to follow different evolutionary paths in the X-ray band. The powerful FSRQs evolve positively (i.e., there is an increasing number of sources at higher redshift, and/or the sources' luminosity increases with redshift) up to redshift $z>4$ \citep[e.g.][]{Ajello_2009, Marcotulli_2022}. BL Lac sources, fainter and for which redshifts are often tenuous, appear not to evolve, or to evolve negatively in the X-ray band \citep{Beckmann_2003,Ajello_2009}. 

The latest blazar evolutionary model in the hard X-ray band has confirmed that the peak of the evolution of the most luminous FSRQs ($L_X>10^{47}\,\rm erg~s^{-1}$) lies beyond redshift 4 \citep[$z_{\rm peak}=4.3\pm0.5$][]{Marcotulli_2022}. These sources are powered by the most massive black holes (see Section~\ref{sec:mev_bla}), and hence closely trace the evolution of the most massive black holes in the Universe. However, owing to the sensitivity limits of current X-ray surveys \citep[e.g.,][]{BAT_105}, we have yet to detect a turnover in the luminosity function at the lowest X-ray luminosities. This break, expected both as a result of beaming \citep{Urry_Shafer_1984} and due to the preferred evolutionary path, hampers the determination of how these sources evolve through cosmic time, i.e., whether there are more (density evolution) or more luminous (luminosity evolution) sources at earlier times. 
If we could tell the two evolution scenarios apart, this would have a profound impact on our current
understanding of SMBH evolution. 
Moreover, studies based on radio-selected samples, found that blazar evolution follows the trend observed in the overall AGN population with a density peak $z\sim2-3$ \citep{Caccianiga_2019,Ighina_2021, Diana_2022}. Only by finding the bulk of the X-ray blazar population $z>4$ would enable us to confirm the location of the peak, which in turn will inform us on the SMBH space density at those distances.

Using the latest hard X-ray luminosity function, we calculated the number of blazars detectable at different redshift bins in a 1 deg$^2$ survey, as planned for the HEX-P mission \citep{Civano23} with the predicted source sensitivity in the $8-24\,\rm keV$ band for the HET instruments. The right panel of Figure~\ref{fig:logn_logs} shows these predictions based on the luminosity (pink hatched boxes) and density evolution model (green solid boxes). As can be seen, depending on the chosen evolution model, the number of sources detected at different redshifts is significantly different, with the luminosity evolution prediction extending up to $z>8$. 
With its sensitivity, HEX-P would be able to answer these questions: are FSRQs evolving in density or luminosity? and where exactly is the peak of the population ($z_{\rm peak}\pm0.1$)? We note that blazars, or possibly nascent jets, have been already detected beyond $z=6$ \citep[e.g.,][]{Zhu_2019,Belladitta_2020}. The left panel of Figure~\ref{fig:logn_logs} shows the predicted source count distribution ($\log N-\log S$; i.e. number of sources as a function of energy flux) of X-ray blazars in the $8-24\,\rm keV$ band as extrapolated from the X-ray best-fit luminosity evolution function in the same 1 deg$^2$ survey. With the capabilities of the HET, we will finally be able to detect the break in {$\log N-\log S$} and break the degeneracy that would lead us to understand for the very first time the type of evolution taking place in this source class. 

Moreover, these sources, though contributing only $\sim2-10\%$ to the cosmic X-ray background above $14\,\rm keV$, are predicted to be the major contributors to the MeV background. Assuming a luminosity evolution scenario, estimates are of the order of 70\%, while for the density evolution, they are up to 100\% \citep[][]{Marcotulli_2022}. However, such numbers come from extrapolations of the best-fit X-ray luminosity function to the MeV band, assuming an average high-energy (X-to-$\gamma$-ray) SED of the MeV blazars class derived from time-averaged Swift/BAT and Fermi-LAT data. As shown in \citet{2020ApJ...889..164M}, quasi-simultaneous soft (Swift/XRT) and hard (NuSTAR) X-ray data, combined with $\gamma$-ray ones, are necessary to constrain the location (both in flux and energy) of the high-energy SED peak, translating constraints to the contribution of this class to the background. Besides allowing us to detect the bulk of the high-redshift population, HEX-P, with its lower background at hard X-rays, will allow probing the high-energy SED very close to its peak ($E>50\,\rm keV$) pinpointing its location and the shape of the underlying electron distribution. 
This, together with the knowledge of the preferred evolution model, will provide tighter constraints on the contribution of MeV blazars to the cosmic MeV background, allowing us to reduce the uncertainties from 30\% to 10\%.

Similarly, the prospects to finally disentangle the preferred evolutionary paths for BL Lacs in the X-ray band are bright with HEX-P. Current photometric and spectroscopic follow-up of $\gamma$-ray detected BL Lacs are increasing the redshift completeness of this challenging source class. A sensitive X-ray telescope such as HEX-P would enable a complete sample of X-ray selected BL Lacs, even to lower luminosity (see Section~\ref{sec:mev_synch_bll}), and determine for the first time their evolution in the full $0.2-80\,\rm keV$ band.

\section{Resolving nearby jets}\label{sec:jet_resolve}

\subsection{Blazars}

Jets of active galaxies can be hundreds to thousands of kiloparsecs long \citep[e.g.,][]{Oei_2022}. The extended structure is most easily observed in the radio band, where the low-energy synchrotron emission can be observed along the entire jet. However, many jets also show extended components (extended jets and/or hotspots along the jet flow) in other energy bands, including the X-ray domain. Observations with Chandra increased the number of detected X-ray jets considerably and rekindled the debate on the X-ray emission processes in extended jets \citep[see][for a review]{Harris_2006}. Especially the explanation of hard X-ray spectra (harder than the radio-through-optical extrapolation) was strongly debated. Two main models emerged: Inverse-Compton scattering of CMB photons by low-energetic electrons \citep[which would produce the radio through optical emission through synchrotron emission;][]{Tavecchio+00} or synchrotron emission of a second, high-energy electron population \citep[e.g.,][]{Hardcastle06,Jester+06}. While the IC/CMB model was favored for some time, several studies over recent years have shown that this is probably incorrect in many sources due to a lack of associated $\gamma$-rays \citep[e.g.,][]{meyer+17,breiding+17}. 

For example, an extended X-ray jet has been observed in the flat-spectrum radio quasar 3C~273 \citep{marshall+2001}, revealing a $\sim 10"$ extended structure with knots. No cut-off is observed in the X-ray spectrum. Multi-wavelength data suggest that the X-ray component is inverse-Compton emission \citep{uchiyama+2006} or a second synchrotron component.

However, in a notable counterexample, the synchrotron explanation probably does not work for the blazar AP~Librae \citep{ZachariasWagner16,Roychowdhury+22}, one of the few BL Lac objects with a resolved, $10"$ long X-ray jet \citep{kaufmann+2013}. The shape of the SED requires at least two emission zones. In fact, the VHE $\gamma$-ray emission observed with H.E.S.S. \citep{2015A&A...573A..31H} has been interpreted as inverse-Compton scattering of dust emission from the inner $1\,$kpc of the jet \citep{Roychowdhury+22}. 

A recent study by \cite{Reddy+23} on jet knots revealed that in most knots the X-ray emission peak is located upstream of the radio emission peak. This might suggest a cooling effect on the particle flow across a shock pointing towards the synchrotron process in both energy bands, as the IC/CMB model requires a one-zone interpretation, where the fluxes should peak at the same location. 
Hence, the debate continues. Observations with HEX-P will be crucial for studies of the hard X-ray emission of nearby, resolved blazar jets. The X-ray spectra do not show any spectral feature; that is, they are described well with a power-law \citep[e.g.,][]{marshall+2001,kaufmann+2013}. This suggests efficient particle acceleration in large-scale jets potentially up to the synchrotron-burnoff limit \citep{deJager+1996}. Unfortunately, the spectral range of the synchrotron-burnoff limit ($50\,$MeV modulo beaming) is currently not well accessible, as the \textit{Fermi}-LAT instrument cannot resolve extended blazar jets \citep[e.g.,][]{breiding+17} in order to derive the spectrum at these energies. The observation of a break or a cut-off in the hard X-ray domain would provide important information on the acceleration and radiation processes inside the extended jets. This would constrain both synchrotron and inverse-Compton models. 

\begin{figure*}
    \centering
    \includegraphics[width=0.45\textwidth]{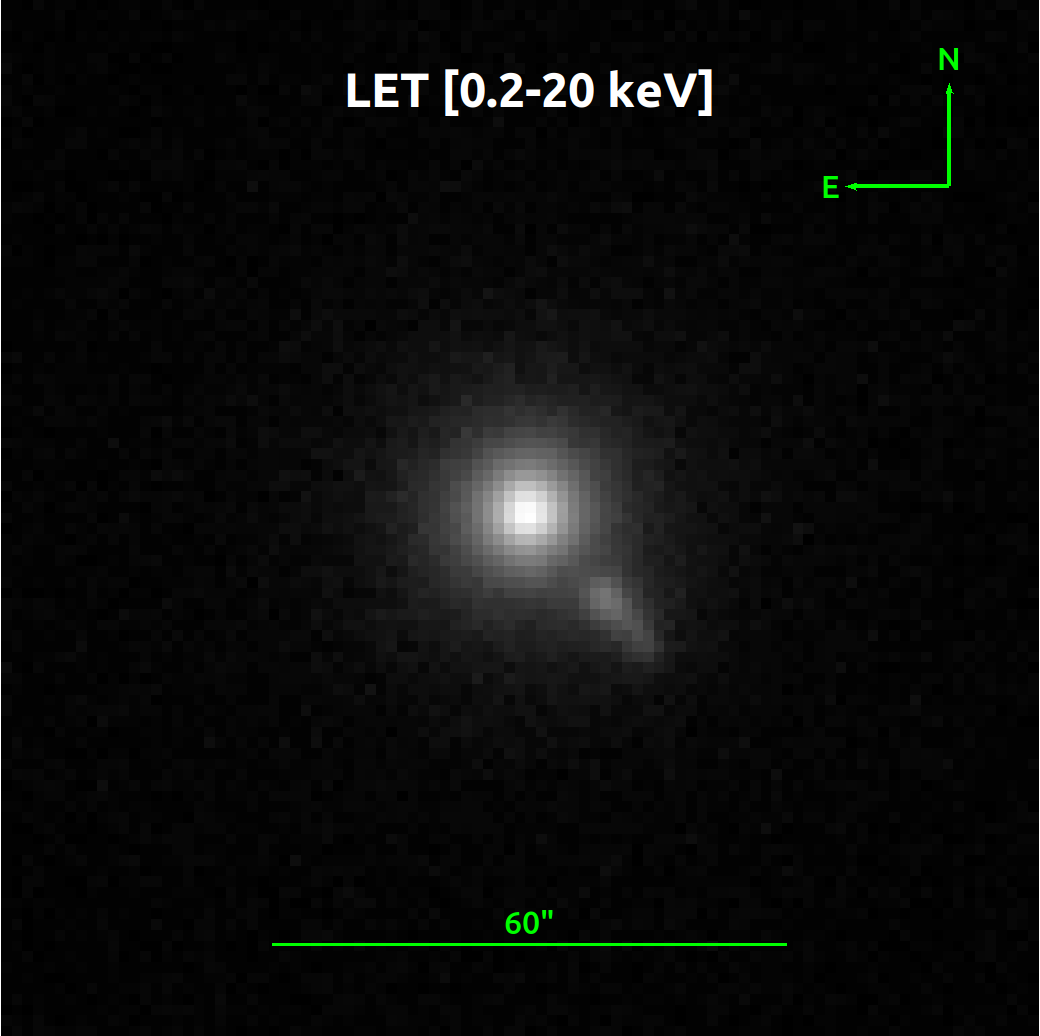}
    \includegraphics[width=0.45\textwidth]{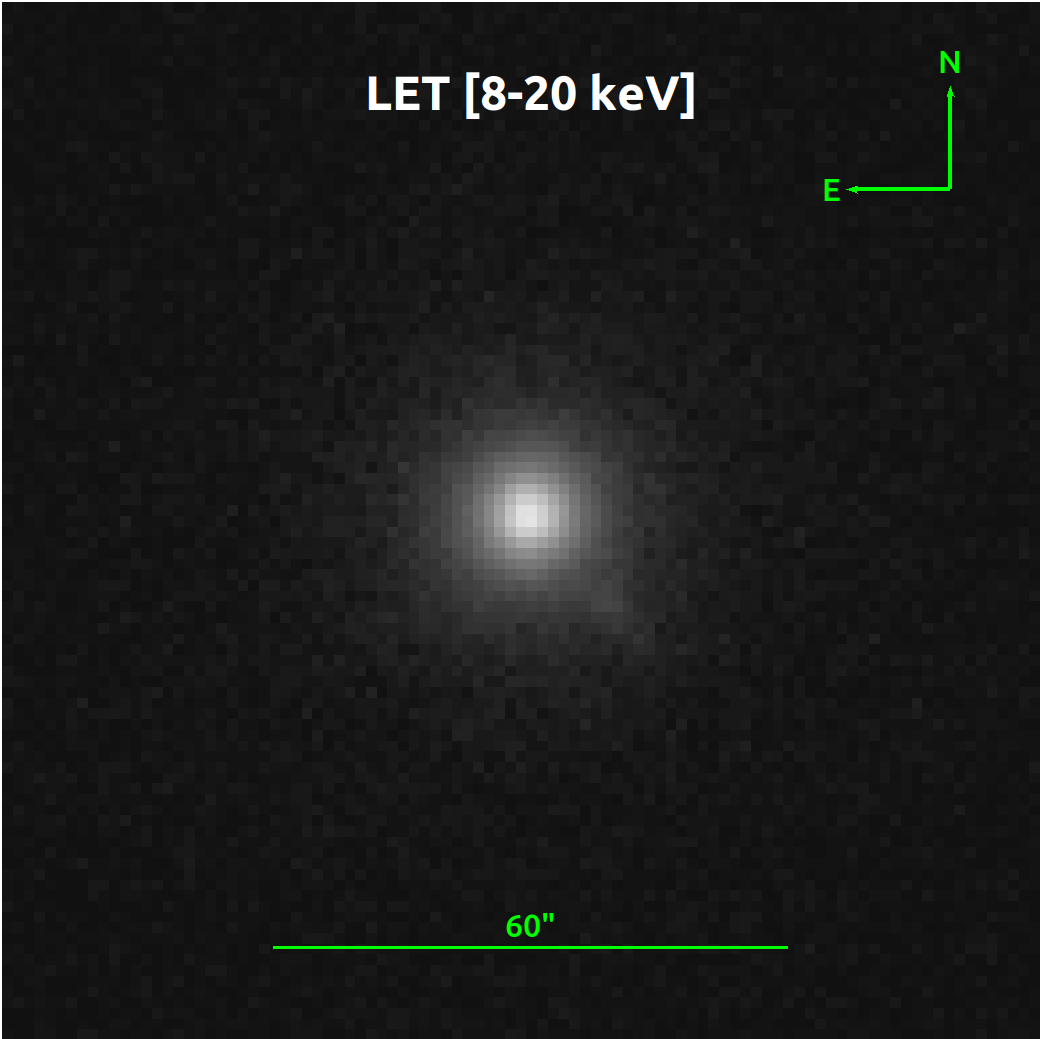}
    \caption{SIXTE simulations of the blazar 3C 273. {\bf Left:} 100 ks observation in the full $0.2-20\,\rm keV$ range with the LET. {\bf Right:} 100 ks observation in the $8-20\,\rm keV$ range with the LET. The field of view in both images is $2' \times 2'$ and we have over-plotted a green line of length $60"$ for scale. We note that $60"$ is also the PSF of NuSTAR; as can been seen, HEX-P imaging capabilities would be key to resolve blazar jets for the first time $>10\,\rm keV$.}
    \label{fig:3c273_ext}
\end{figure*}

Using the 47 sources labeled as \textit{BL Lac} or \textit{CDQ} in the Atlas-X\footnote{\url{https://astro.umbc.edu/Atlas-X/}} \citep{Reddy+23} and the XJET\footnote{\url{https://hea-www.cfa.harvard.edu/XJET/}} catalogs (most of which are blazars), we determine that HEX-P's angular resolution of $\leq 5"$ is sufficient to resolve up to 20 known (blazar) jets at hard X-rays. 6 sources have jets longer than $10"$ including, for example, 3C~273  (cf., Fig.~\ref{fig:3c273_ext}), 9 sources range between $5"$ and $10"$, while 5 jets are about $5"$ long which might still be detectable. Thus, HEX-P will help determining the spectral features (breaks or cut-offs) at X-rays beyond $10\,$keV, which are so far unknown.  In turn, the acceleration processes in extended jets will be strongly constraint, which will significantly enhance the modeling of these extended structures \citep{2015MNRAS.453.4070P,2019MNRAS.482.4798L,2022MNRAS.512.3948Z}, and which would greatly enhance the theoretical understanding of jets as a whole.

\subsection{TeV-detected Radio Galaxies} 
Radio galaxies are typically non-aligned AGN with significantly reduced Doppler beaming. Depending on the X-ray production process, their extended X-ray jets may still be observable, as beaming is not as important depending on the particle energies in the jet. 
Here, we discuss a special class of radio galaxies; namely those, which are detected at TeV energies. So far, four bona-fide radio galaxies are listed in the TeVCAT,\footnote{\url{http://tevcat2.uchicago.edu/}} of which three have extended X-ray jets: M~87 \citep{2006Sci...314.1424A,2009Sci...325..444A,2012ApJ...746..151A}, Centaurus~A \citep{2009ApJ...695L..40A,HESS20}, and 3C~264 \citep[however, the X-ray jet is less than $5"$ long]{2020ApJ...896...41A}. Centaurus~A is a special case, as it is not only the closest AGN to Earth, but also, so far, the only AGN where the TeV emission is spatially resolved. The TeV emission in Centaurus~A is extended along the axis of the X-ray and radio jet \citep{HESS20}.
This could indicate the presence of high-energy particles on kpc scales that are inverse Compton scattering dust emission or starlight, or it could be associated to synchrotron radiation from ultra-high energy electrons \citep[e.g.,][]{Wang_2021}.

\begin{figure*}
    \centering
    \includegraphics[width=0.325\textwidth]{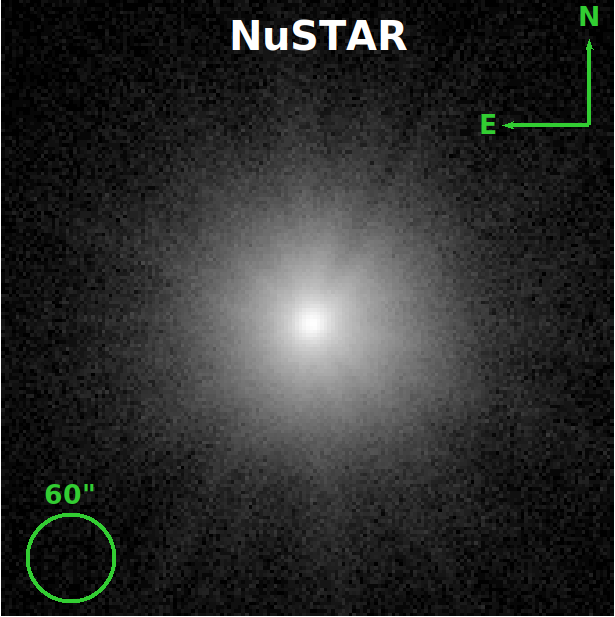}
    \includegraphics[width=0.323\textwidth]{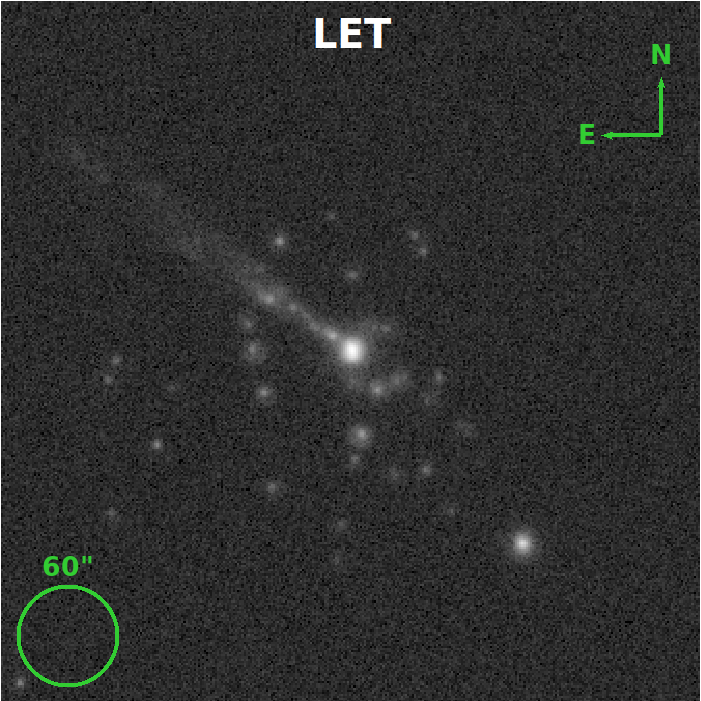}
    \includegraphics[width=0.324\textwidth]{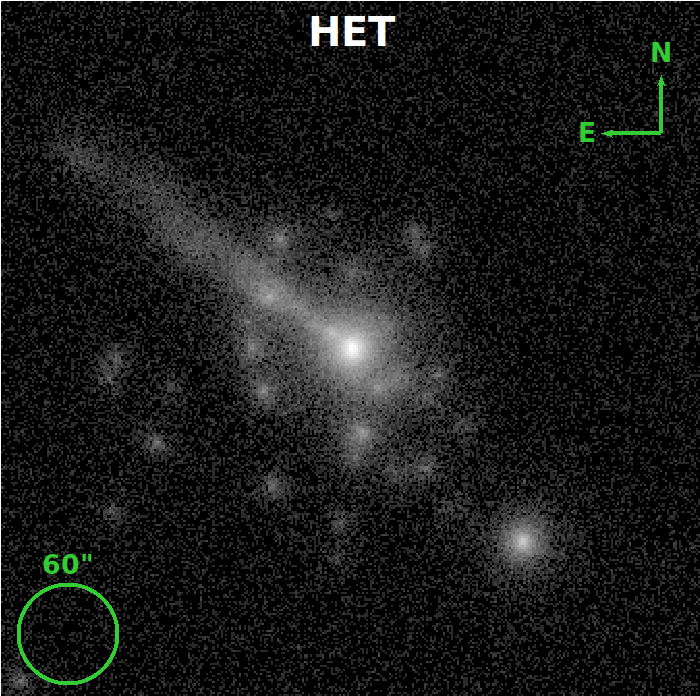}
    \caption{{\bf Left:} NuSTAR FPMA $3-80\,\rm keV$ image of the radio galaxy Centaurus A (obsid: 60001081002). {\bf Middle \& Right:} SIXTE simulations of the same galaxy with HEX-P based on the Chandra image of the source ({\it Middle:} 500 ks observation in the $8-20\,\rm keV$ range with the LET; {\it Right:} 500 ks observation in the $20-50\,\rm keV$ range with the HET). HEX-P would enable us, for the first time, to resolve extended jets in the hard X-ray band as well as study point sources visible in the Chandra image at soft X-rays (see Lehmer et al., in prep). The images are all $7' \times 7'$ and we have over-plotted a green circle of diameter $1'$ for scale.}
    \label{fig:cenA_ext}
\end{figure*}

Observations with HEX-P of these extended jets, coupled with the knowledge that these jets emit TeV $\gamma$ rays at large distances from the black hole, will be crucial for constraining the location and physics of the $\gamma$-ray production. As mentioned above, cut offs or breaks in the X-ray spectra would reveal maximum particle energies (or at least the acceleration efficiency) indicating that such a particle population would probably not produce $\gamma$ rays. Determining spatially resolved spectra of nearby jets (such as Centaurus~A; see Fig.~\ref{fig:cenA_ext}) would allow us to pinpoint the regions where the extended $\gamma$-ray emission is produced. 

Along with HEX-P and its broad X-ray coverage, multiwavelength observations with future high-angular-resolution instruments will be crucial to unlock the secrets of these objects. Most notably, collaborative observations by HEX-P with CTAO will be important to constrain the jet morphology and processes.

\section{Conclusions}\label{sec:concl}

The HEX-P mission offers several key strengths for studying jetted AGN. Its broad-band sensitivity, spanning from 0.2 to 80 keV, provides deep coverage in relatively short exposure times. Additionally, HEX-P exhibits excellent angular resolution at high energies and the unique capability to point in the antisolar direction. These remarkable capabilities set HEX-P apart from both current and currently planned X-ray observatories, making it an unparalleled asset for the study of blazars in the next decade.
In particular:
\begin{itemize}

\item HEX-P's potential extends to time-domain and multi-messenger studies.
HEX-P has the capability to independently discover the expected signatures of hadronic accelerators. In Section \ref{sec:particles}, we illustrated how HEX-P can detect spectral changes during an X-ray flare 
as expected from the presence of a hadronic component. Moreover, by observing blazars that are detected by IceCube, HEX-P can identify
the X-ray signatures expected from hadronic sources. In Section \ref{sec:mm}, we demonstrated that HEX-P can detect the signature of proton-induced cascades associated with neutrino emission. It accomplishes this by distinguishing between power-law emission, typical of a leptonic synchrotron component, from broken power laws, expected in the case of proton-induced cascades.
Note, however, that these signatures are not unique to hadronic processes and may also arise in multi-component leptonic emission scenarios. \\

\item The deep broad-band sensitivity of HEX-P will enable the study of rapid variability on minute timescales. As shown in Section \ref{sec:variability}, these investigations provide insight into the connection between the acceleration and cooling of charged particles. Previously, such studies were limited to only a few objects over an energy band spanning less than one decade. HEX-P will expand this research to a significantly larger number of objects and offer an energy band spanning more than two decades.
Furthermore, HEX-P's ability to pinpoint extremely fast variability, comparable to the light-crossing time of the system, will reveal invaluable information about particle acceleration mechanisms. In Section \ref{sec:variability}, we demonstrated how HEX-P can measure the lightcurve of a fast flaring blazar (e.g., Mrk 421) at both low ($<$10 keV) and high ($>$10 keV) X-ray energies, thereby identifying the signature of magnetic reconnection.\\

\item Section \ref{sec:source_classes} showed how HEX-P will allow investigation of the accretion-jet connection by measuring the average X-ray SED of selected blazars and constrain variation in the SED linked to possible accretion fluctuations. Moreover, HEX-P will detect and study MeV blazars (the most luminous type of blazars) well above the current $z\approx3$ limit and possibly all the way to $z=6$. \\

\item As shown in Section \ref{sec:jet_resolve}, for the first time above 20\,keV, HEX-P will be able to resolve jets of blazars and radio galaxies, thus pinpointing the regions where the extended emission is produced. 
HEX-P will be able to discriminate between the emission of the core and that of the jet allowing to constrain potential spectral cut-offs in the jet. This would yield an important constraints on the maximum energy of the particle population.

\end{itemize}

Lastly, HEX-P exhibits strong synergies with current and future planned facilities, including CTAO, IceCube, IXPE, COSI, and the current and future CMB facilities (Section~\ref{sec:synergies}). These collaborations enable the study of extremely fast flaring blazars, the detection and study of multi-messenger sources, and the investigation of hadronic emission components in blazars.  An early 2030s launch of HEX-P will significantly enhance the science provided by those facilities and answer many key questions about the most powerful jets in the Universe.

\section*{Author Contributions}
L.M.\ led the creation of the manuscript, authoring the Abstract, Section~\ref{sec:intro}-\ref{sec:simulations} and Section~\ref{sec:blazar_evolution}. They were responsible for producing the Figures~\ref{fig:peak_shift} and \ref{fig:reconnection} based on the theoretical simulations performed by H.Z. They produced the simulations and SED fitting for Figure~\ref{fig:masq_bll}-\ref{fig:logn_logs}, and produced the images simulations shown in Figure~\ref{fig:3c273_ext}-\ref{fig:cenA_ext}. \\
M.A. authored Section~\ref{sec:masquerading} and Section~\ref{sec:concl}. \\
G.M. authored Section~\ref{sec:lep_mod} and Section~\ref{sec:spt}, and rendered Figure~\ref{fig:pks2155_sed}.\\
M.P., M.B. and H.Z. authored Section~\ref{sec:modeling_hadro_lepto} and performed the theoretical simulations for Figure~\ref{fig:aplib} and \ref{fig:aplib-HR}.\\
G.V. conducted the simulations showed in Figure~\ref{fig:aplib} and \ref{fig:aplib-HR} based on M.P., M.B. and H.Z. simulations. \\
F. M. authored Section~\ref{sec:mm}.\\
A.G., L.C. and H.Z. authored Section~\ref{sec:variability}\\ 
L.C. authored Section~\ref{sec:CTA} and Section~\ref{sec:extreme_bllac}, and provided the modeling and simulation shown in Figure~\ref{fig:Cascades} and \ref{fig:extremes}.\\
I.L. and R.M. co-authored Section~\ref{sec:ixpe}.\\
T.S. authored Section~\ref{sec:accre_fl} and created Figure~\ref{fig:fsrq_sed}.\\
A.G. produced Section~\ref{sec:mev_bla}.\\
L.C. and M.E. authored Section~\ref{sec:extreme_bllac}.\\
M. Z. authored Section~\ref{sec:jet_resolve}.\\
K. M. provided the differential sensitivity curves shown in Figure~\ref{fig:peak_shift} and \ref{fig:fsrq_sed}. \\
All authors contributed to the discussion on the different science cases and the final editing of the manuscript.

\section*{Funding}

The work of D.S.\ was carried out at the Jet Propulsion Laboratory, California Institute of Technology, under a contract with NASA.


\newpage


\bibliography{report}   
\bibliographystyle{Frontiers-Harvard}   


\end{document}